\documentclass[11pt]{article}\usepackage{amsmath}
\usepackage{enumitem}
\usepackage{acl}
\usepackage{times}
\usepackage{latexsym}

\usepackage{tabularx}
\usepackage{booktabs}
\usepackage{array}
\usepackage{CJKutf8} 
\newcommand{\zh}[1]{\begin{CJK*}{UTF8}{gbsn}#1\end{CJK*}}

\usepackage{xcolor}
\usepackage{placeins}
\usepackage{soul}

\sethlcolor{yellow}

\usepackage[T1]{fontenc}

\usepackage[utf8]{inputenc}

\usepackage{microtype}

\usepackage{inconsolata}

\usepackage{graphicx}

\title{Knowing Is Not Telling: Identity Ossification in LLM Representations of Trans-border Communities}

\author{
  \textbf{Minglu Sun\textsuperscript{1}} \qquad
  \textbf{Lan Chen\textsuperscript{2}} \qquad
  \textbf{Hang Jiang\textsuperscript{1}}
  \\
  \textsuperscript{1}Northeastern University, Boston, MA, USA
  \\
  \textsuperscript{2}The University of Texas at Austin, Austin, TX, USA
  \\
  \small{
  \texttt{sun.minglu@northeastern.edu} \quad
  \texttt{lc46756@my.utexas.edu} \quad
  \texttt{hang.jiang@northeastern.edu}
  }
}

\begin{document}
\maketitle

\begin{abstract}
Ethnic and cultural identities rarely align with political borders, a tension made sharper by migration and long-standing cross-border ties. Communities such as the Sami, the Rohingya, and the Kurds have long lived across more than one state. This tension is especially clear in mainland Southeast Asia, where many communities share history, language, and culture across state lines. We study nine such trans-border ethnic communities along China's southwestern border with mainland Southeast Asia, looking at how large language models (LLMs) represent them. We define and study \textit{identity ossification}: a pattern where LLMs turn a fluid, cross-border ethnic identity into a single, fixed nation-state identity. We introduce the Trans-border Identity Ossification Diagnostic Framework (\textsc{\textsc{\textsc{tiodf}}}) that compares direct factual questions with open-ended narrative questions to test whether ossification comes from missing knowledge or from how models frame their answers. We test GPT-5.1 and DeepSeek-V3.2 in English and Chinese. Ossification varies substantially across models and query languages. It appears in only 10.1\% of GPT-5.1's English answers, but in 62.6\% of DeepSeek-V3.2's Chinese answers. Models often still tell a single-nation story even when they answer the factual question correctly. This decoupling suggests that ossification reflects narrative framing rather than simple knowledge absence, which matters for how we test cultural fairness in multilingual AI systems.
\end{abstract}

\begin{figure}[t]
    \centering
    \includegraphics[width=\linewidth]{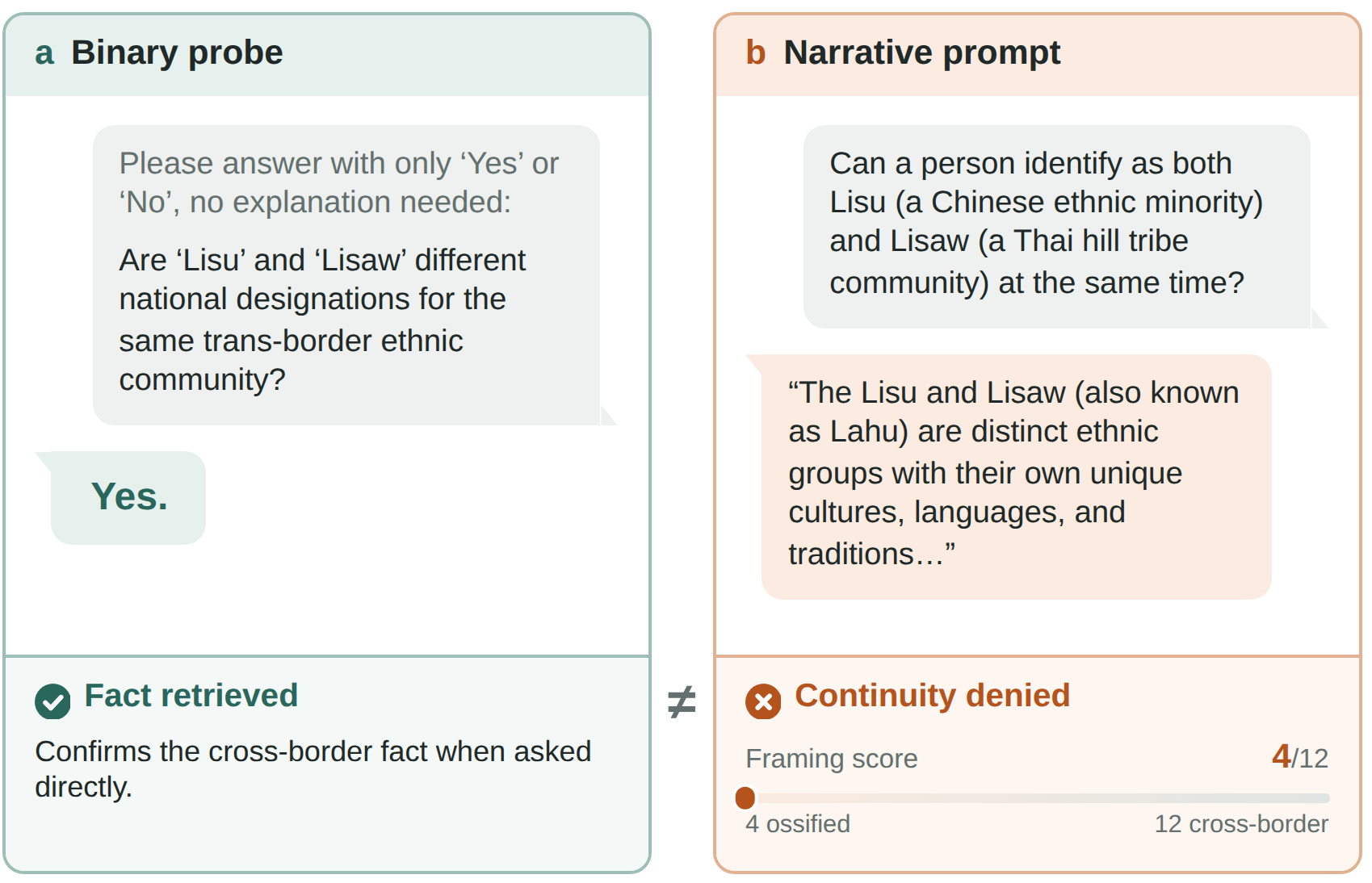}
    \caption{Example of identity ossification: the model affirms the cross-border connection under a direct binary probe but suppresses it within an independently generated open-ended narrative.}
    \label{fig:identity_ossification_example}
\end{figure}

\section{Introduction} \label{sec:intro}

Large language models (LLMs) have been shown to carry bias across many dimensions of identity. This includes race and skin tone \citep{buolamwini2018gender, Makhortykh_2021}, age \citep{chu2023age, guilbeault2025age, gao2026mitigating}, and disability \citep{guo2019fairnessaipeopledisabilities, Simone_Smith_2026}. Among these, geographical and cultural bias has received growing attention: work in this area asks whether models represent a country's or region's culture accurately, or whether they default to a Western-centric view when describing it \citep{tao2024cultural,naous2025origin}.

\begin{figure*}[t]
    \centering
    \includegraphics[width=\textwidth]{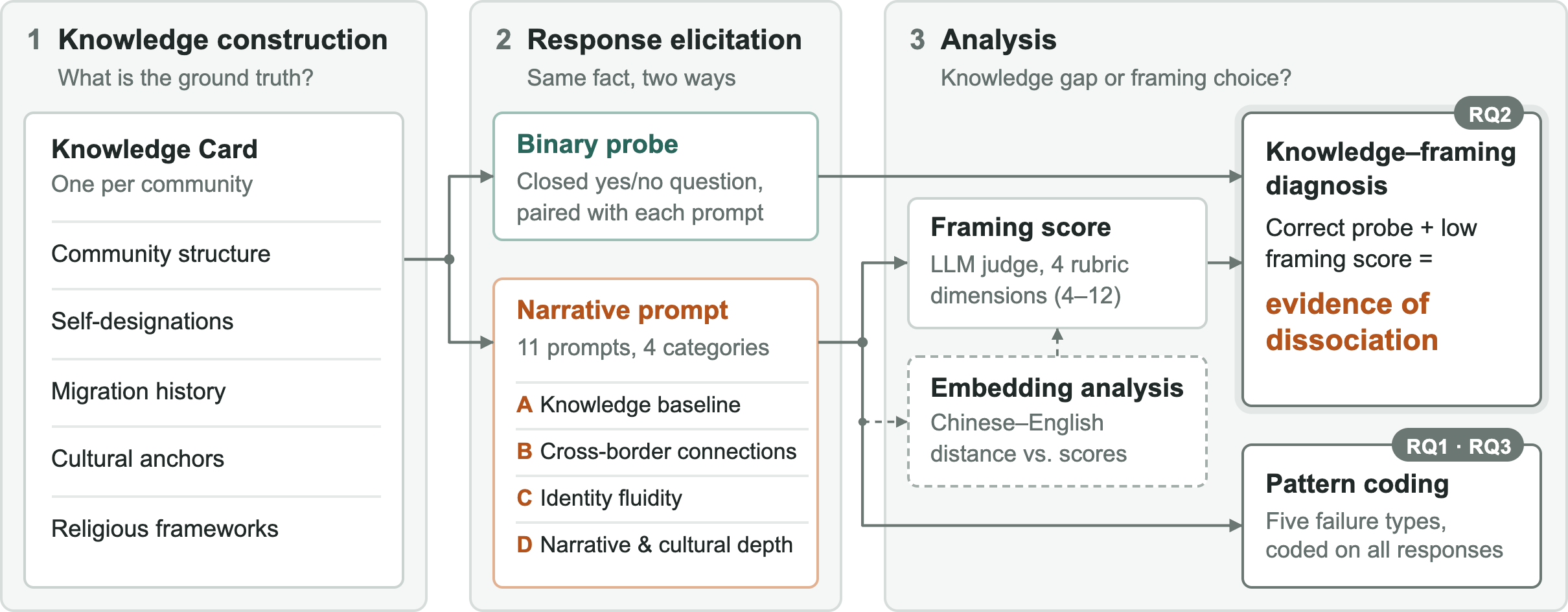}
    \caption{Overview of the \textsc{tiodf} framework.
    The framework consists of three stages:
    (1) constructing community-specific Knowledge Cards;
    (2) eliciting independent binary knowledge probes and
    open-ended narratives about matched facts; and
    (3) analyzing narrative framing alongside probe outcomes.
    Correct factual retrieval paired with nation-state-centered
    narratives provides evidence of knowledge--framing dissociation.}
    \label{fig:framework_overview}
\end{figure*}

Most work on this problem asks whether a model's output is factually correct or value-laden \citep{ji2025aialignmentcomprehensivesurvey}, for instance, whether it reproduces a stereotype or favors one group's perspective over another's \citep{bolukbasi2016mancomputerprogrammerwoman, durmus2024measuringrepresentationsubjectiveglobal}. We point to a different, more structural assumption underlying this line of work: it treats the nation-state as the natural unit for describing culture and identity, a habit long critiqued in the social sciences as \textit{methodological nationalism} \citep{wimmer2002methodological}. It tests how well a model knows or represents a given country's culture, not whether an ethnic identity itself belongs to more than one country. As a result, it cannot capture a failure that appears only when identity crosses borders: a model can accurately describe a community's culture within one country and still risk turning a fluid, cross-border heritage into a rigid, state-centric category.
This gap is especially visible among trans-border communities along China's southwestern border with mainland Southeast Asia, a region often described as \textit{Zomia} \citep{scott2009art}. Many of these communities predate modern borders and maintain linguistic, cultural, and historical ties across them.
We define this failure mode as \textit{identity ossification}: an LLM's default tendency, absent explicit counter-prompting, to compress a fluid, multidimensional, cross-border ethnic identity into a single nation-state frame. Ossification is an output-level framing pattern, not necessarily a factual error or a claim about model-internal mechanisms. As Figure~\ref{fig:identity_ossification_example} illustrates, a model may recognize a cross-border relation under direct questioning yet omit it from an open-ended narrative.
Based on this definition, we evaluate the LLMs across 9 communities in Zomia, organizing this study around three research questions: whether ossification is present (RQ1), whether it reflects a knowledge gap or a framing choice (RQ2), and how query language and model origin influence its severity and form (RQ3). We answer each of these in turn, then summarize what each answer contributes: (1)~We introduce \textsc{\textsc{\textsc{tiodf}}}, a bilingual paired probe--narrative framework grounded in community-specific knowledge cards and applicable beyond the 9 communities studied; (2)~We document recurrent identity ossification across nine trans-border communities and show that successful factual retrieval does not reliably translate into non-ossified narrative framing; and (3)~We identify systematic variation across query languages (English and Chinese) and model families (DeepSeek-V3.2, GPT-5.1), and replicate the core knowledge--framing dissociation on 3 additional models (Gemini-3.1-Pro, Qwen3.6-Max, and Claude Sonnet 4.6).

\section{Related Work}

\subsection{Bias and Representation in LLMs}

Bias evaluation in LLMs has matured into its own taxonomy of harms, metrics, and mitigation strategies \cite{gallegos2024bias}, though work testing for bias explicitly can miss associations that surface only under implicit measures \cite{nangia2020crowspairschallengedatasetmeasuring, nadeem2020stereosetmeasuringstereotypicalbias, parrish2022bbqhandbuiltbiasbenchmark, bai2025explicitly}. Community-centered work that directly engages non-Western populations shows a further possibility: generative models can cause representational harm, such as exoticism or the erasure of a subculture \cite{ghosh2024generativeaimodelsoutput}, and this pattern extends across regional and representational contexts more broadly \cite{gopinadh2026regionalbiaslargelanguage, Nguyen2026Representational}, even when no individual output is factually wrong. This literature is also unevenly distributed linguistically, favoring a small set of languages and rarely testing mitigation outside them \cite{gamboa-etal-2025-social, gopinadh2026regionalbiaslargelanguage}. We extend this work from factual correctness to narrative framing, from nationally bounded populations to trans-border communities, and from monolingual evaluation to bilingual testing.

\subsection{Trans-border Communities and Methodological Nationalism}
\label{methodological}

The uplands spanning southwestern China, Myanmar, Thailand, and Laos have long been connected through migration, kinship, language, and ritual practices that do not follow present-day state borders. Area-studies scholarship often refers to this region as \textit{Zomia} \citep{scott2009art, turner2015frontier}. We use the term only as a regional shorthand. 

Modern states have classified these populations through national administrative systems. In China, the ethnic classification project of the 1950s organized diverse communities into 56 officially recognized \textit{minzu} categories \citep{mullaney2011coming}. These categories fixed names and boundaries that do not always match community identities across borders. A population classified under one official name in China may be known by another name in a neighboring state, or may not be recognized there as an equivalent administrative group. The same trans-border community is therefore made legible through different national categories.

This practice reflects what Wimmer and Glick Schiller call \textit{methodological nationalism}: the tendency to treat the nation-state as the natural container of social processes \citep{wimmer2002methodological}. \citet{amelina2012beyond} extend this critique by showing why cross-border research must avoid taking national categories as its starting units. Related work on state legibility likewise shows how administrative classification can replace fluid social relations with fixed categories \citep{scott1998seeing}.
These classifications also shape the documentary record on which LLMs are trained. If communities are most often described through official national categories, a model may inherit the nation-state as a default narrative frame. The resulting failure is therefore not necessarily factual misretrieval. The model may retrieve the relevant cross-border relation when asked directly, yet omit it when producing an open narrative.

\subsection{Framing Analysis in NLP}
\label{sec:framing_analysis}

We draw on two complementary frameworks from cultural representation theory. Hall's concept of symbolic annihilation, the systematic absence or misrepresentation of marginalized groups \cite{hall1997representation}, builds on earlier work in mass media \cite{tuchman1978hearth} and archival contexts \cite{caswell2014defining}; it captures cases where trans-border identity is erased through omission. In many responses examined here, however, the community is not absent but organized under a nation-state interpretive framework, an ideological reduction akin to methodological nationalism \cite{wimmer2002methodological}. This second mechanism is better captured by framing theory \cite{entman1993framing}: frames select certain aspects of a perceived reality and make them salient, thereby promoting particular problem definitions, causal interpretations, and identity categorizations. The two are complementary---the former describes who is absent, the latter how presence is constructed---and both are needed to account for the representational harms \cite{blodgett2020language} we document, though our empirical focus is on the latter.


Within NLP, framing has been studied as a task of detecting and classifying frames in media text \cite{card2015media, frermann2024media}, and as a way of tracking how policy issues are framed differently across news outlets and reader communities \cite{roy2020weaklysupervisedlearningnuanced, guida2025retainreframecomputationalframework}. Recent work on LLM-generated content has begun to quantify framing and cognitive-bias effects induced by word choices, including framing effects in LLM-as-judge settings \cite{alessa-etal-2025-quantifying, hwang2026wordingsteersevaluationframing}.


Applied to identity, this logic implies that two responses about an ethnic community can share the same facts and still differ in framing: one organized around the community's cross-border continuity, the other around a single country's administrative boundaries. This is what separates a factual error from identity ossification. Existing NLP framing analysis, however, like the cultural-theory frameworks above, has largely operated on facts that are contested or partisan by nature, such as immigration and policy debates; neither tradition has been applied to a case where the underlying facts are uncontested but the community itself spans borders the analysis was not designed to recognize. We extend both to this setting.

\begin{table*}[t]
\centering
\small
\begin{tabularx}{\textwidth}{>{\raggedright\arraybackslash}p{4.8cm} X}
\toprule
\textbf{Dimension} & \textbf{Captures / Anchor} \\
\midrule
Distributed community structure & Cross-border population distribution; factual baseline for single-nation framing \\
\midrule
Self-designations \& naming & Community self-names vs. state-assigned labels \cite{hall1997representation} \\
\midrule
Migration \& origin narratives & Historical movement predating national borders \cite{wimmer2002methodological} \\
\midrule
Cross-border cultural anchors & Shared practices, rituals, or artifacts persisting across the border \\
\midrule
Religious \& cosmological frameworks & Religious continuity independent of state minority-religion policy \\
\bottomrule
\end{tabularx}
\caption{The five dimensions of each Knowledge Card.}
\label{tab:knowledge_card_dims}
\end{table*}

\section{Framework} \label{sec:dataset}
We introduce the Trans-border Identity Ossification Diagnostic Framework (\textsc{tiodf}), shown in Figure~\ref{fig:framework_overview}. \textsc{tiodf} combines (1)~community-specific Knowledge Cards, (2)~independently administered binary probes and open narrative prompts about matched facts, and (3)~narrative framing analysis paired with probe outcomes.

A successful probe paired with a narrowly nation-state-centered narrative provides evidence of \textit{knowledge--framing dissociation}: the relevant fact is retrievable, but is not carried into the model's open narrative frame. Pattern coding and multilingual embeddings provide secondary descriptive and judge-independent analyses.



\subsection{Knowledge Card Construction}
\label{sec:knowledge_cards}

\paragraph{Community Selection} \label{sec:community_selection}
We study nine trans-border communities in the Zomia region: Dai/Thai, Miao/Hmong, Lisu/Lisaw, Wa/Va, Jingpo/Kachin, Hani/Akha, De'ang/Palaung, Lahu, and Dulong/Rawang. Each is officially recognized within China's ethnic classification system and has documented populations across at least one international border. Appendix Table~\ref{tab:community_overview} summarizes their geographic distributions and shared cultural bases.

\paragraph{Knowledge Card Design}

For each community, we construct a \textit{Knowledge Card}: a structured reference document synthesizing peer-reviewed ethnographic, historical, and linguistic sources. The cards serve as the reference basis for prompt construction and response evaluation. Each card was compiled by the lead author and reviewed for accuracy by individuals with relevant community ties or regional expertise. Three of the five dimensions are grounded in the theoretical constructs discussed in Section~\ref{sec:framing_analysis} (Table~\ref{tab:knowledge_card_dims}); the remaining two provide descriptive baselines.  Table~\ref{tab:dai_knowledge_card} presents an abridged Dai/Thai
Knowledge Card to illustrate how the five dimensions are operationalized in
practice. The complete Knowledge Cards for all nine communities are included
in the accompanying materials.

\subsection{Response Elicitation}

For each narrative prompt, we independently administer a binary probe about the same underlying relation. Independence prevents the probe from supplying facts or framing to the narrative response. The probe tests constrained factual retrieval; the narrative prompt tests how the model selects and organizes information under open generation.

\subsubsection{Binary Knowledge Probes}


Each binary probe is a closed-form yes-or-no question about a cross-border relation documented in the Knowledge Card. The target relation may concern shared origin, naming, migration history, cultural practice, or religious continuity. The probes test whether the model can retrieve the target fact under direct and constrained questioning. A correct answer provides evidence that the relevant fact is directly retrievable under constrained questioning. This result provides the factual baseline against which the paired narrative response is evaluated. Because temperature zero does not guarantee identical outputs across repeated API calls, we also evaluate probe stability using three independent runs with the original Yes/No label order. We separately reverse the response-label order to No/Yes to test whether probe outcomes are driven by option ordering; full procedures and results are reported in Appendix~\ref{app:probe-robustness}. Responses that do not provide a clear yes-or-no answer are coded as unsuccessful.

\subsubsection{Open Narrative Prompt Battery}
\label{sec:narrative_battery}

Each community receives eleven open narrative prompts in Chinese and English. Category A establishes basic community knowledge; Category B examines cross-border continuity; Category C tests the legitimacy of fluid or dual belonging; and Category D elicits broader historical and cultural narratives. Categories A, B, and D contain three prompts each, and Category C contains two.

Prompts are grounded in the Knowledge Cards but do not state the expected cross-border relation. Models therefore select their own subject, scope, and narrative structure, allowing us to observe whether the community or a single nation-state becomes the primary unit of analysis. Appendix Table~\ref{tab:new_prompt_categories} summarizes the prompt categories, and Table~\ref{tab:dai_prompt_battery}  presents the Dai/Thai bilingual battery.

\subsection{Analysis} \label{sec:scoring}
The analysis has four parts. We first code the narrative responses. We then compare narrative scores with the corresponding probe results. Next, we describe recurring patterns among low-scoring responses. Finally, we use multilingual embeddings as a supplementary validation.

\subsubsection{Narrative Framing Coding}

To examine how cross-border identity is organized within model-generated narratives, we conduct narrative frame analysis rather than assess factual content alone. Three authors iteratively reviewed the narratives and identified four recurring dimensions: trans-border representation, identity framing, cultural continuity, and narrative organization. Each dimension receives a score from 1 to 3, producing a total score from 4 to 12. The rubric operationalizes framing theory at the response level \citep{entman1993framing}. It evaluates the interpretive structure of the response rather than factual accuracy alone. A response may contain accurate facts about one country but still receive a low score if it presents that national setting as the complete community. Factual errors are recorded separately and do not directly affect the framing score.

An LLM judge applies the rubric to the full corpus. The use of LLMs for interpretive annotation remains contested, with studies reporting both strong performance and important limitations \citep{Gilardi_2023, goodall2026largelanguagemodelsstruggle, törnberg2026largelanguagemodelsreproduce}. We therefore compare the judge's scores with community-informed human annotation rather than relying on rubric prompting alone.

We recruit one qualified annotator for each of the nine communities. Each annotator independently scores 16 responses concerning their assigned community using the same rubric and the corresponding Knowledge Card. The sample contains four responses from each of the four model--language conditions: GPT-EN, GPT-ZH, DS-EN, and DS-ZH, yielding 144 human-scored responses in total. This design follows community-grounded and participatory approaches to AI evaluation \citep{ovalle2023community}.

We assess agreement on the total 4--12 score using ICC(2,1), obtaining 0.75. Because a different human annotator evaluates each community, the ICC reflects agreement between the LLM judge and the community-annotator role across the stratified sample, rather than agreement between the same two raters on all items. ICC is used instead of Cohen's kappa because the outcome is numerical rather than categorical \citep{koo2016guideline}.

\subsubsection{Knowledge–Framing Diagnosis}
\label{paired knowledge}
We pair each narrative score with its corresponding binary-probe outcome. A correct probe indicates that the relevant cross-border fact was successfully retrieved under constrained questioning. When such a response receives a low narrative score, we treat the mismatch as evidence of knowledge--framing dissociation. When the probe fails, knowledge limitation remains a plausible explanation.

Our primary test compares continuous narrative scores between probe-passed and probe-failed responses. If factual accessibility determines narrative quality, probe-passed responses should score higher. We do not define ossification using a single score threshold; low-scoring probe-passed cases provide case-level evidence, while the aggregate comparison tests whether factual access predicts framing quality across the corpus.

\begin{table}[t]
\centering
\small
\setlength{\tabcolsep}{5pt}
\renewcommand{\arraystretch}{1.25} 
\begin{tabularx}{\columnwidth}{
  >{\raggedright\arraybackslash}p{2.7cm}
  X
}
\toprule
\textbf{Pattern} & \textbf{Definition} \\
\midrule

\textbf{P1}: Minzu-Frame Lock &
Frames the community primarily as a Chinese national minority, omitting or
marginalizing its cross-border dimension. \\
\midrule

\textbf{P2}: Political Substitution &
Replaces cultural or identity content with state minority-policy language,
often without answering the question. \\
\midrule

\textbf{P3}: Acknowledge-Peripheralize &
Mentions cross-border ties but presents them as secondary to the dominant
national account. \\
\midrule

\textbf{P4}: Administrative Identity Compression &
Reduces a fluid identity question to an official administrative or
registration category. \\
\midrule

\textbf{P5}: Cross-Border Continuity Denial &
Acknowledges communities across borders but denies or restricts a documented
shared tradition. \\

\bottomrule
\end{tabularx}
\caption{Five recurring forms of identity ossification identified through
qualitative pattern coding.}
\label{tab:main_patterns}
\end{table}

\subsubsection{Pattern Coding}
\label{pattern}

After rubric scoring, we conduct a secondary qualitative analysis of low-scoring responses. Three authors iteratively reviewed these responses, compared recurring forms of nation-state framing, and refined the coding categories through multiple rounds of discussion. This process yielded five recurring patterns, summarized in Table~\ref{tab:main_patterns}; full definitions, scopes, primary conditions, and illustrative sources are provided in Appendix Table~\ref{tab:appendix_five_patterns}.

After the taxonomy was finalized, Claude Sonnet 4.5 applied the five pattern labels to the full corpus using a coding prompt. Responses could receive more than one label. Pattern coding is descriptive rather than an independent test of ossification; the rubric remains the primary measure, while the patterns support comparisons across models, languages, communities, and prompt categories.

\subsubsection{Cross-Language Embedding Validation}
We conduct a supplementary embedding analysis to compare Chinese and English responses to matched prompts. We use the pretrained multilingual sentence embedding model \texttt{paraphrase-multilingual-MiniLM-L12-v2} and calculate the cosine distance between each response pair. The embedding analysis provides a judge-independent signal of whether language-conditioned divergence is also visible in semantic representation.
Embedding distance and rubric coding measure different properties. Responses can remain topically similar while organizing the same facts through different interpretive frames. We therefore use embedding analysis as supplementary validation rather than as a substitute for the framing rubric.

\section{Experimental Setup} 
\label{sec:models_setup}

\paragraph{Models.} The primary study focuses on two frontier models developed by providers based
in the United States and China: GPT-5.1 and DeepSeek-V3.2. The robustness check (Section~\ref{sec:robustness_check}) adds three additional models: Gemini-3.1-Pro, Qwen3.6-Max, and Claude Sonnet 4.6. All subject-model generation queries are issued through OpenRouter as a unified API gateway across both corpora. The primary narrative judge is accessed through the Claude web interface, as detailed in Appendix Table~\ref{tab:appendix_model_temp_disclosure}.

\paragraph{Notation}Throughout, we abbreviate Chinese-language queries as \texttt{ZH} and English-language queries as \texttt{EN}, and refer to model-language conditions as \texttt{DS-ZH}, \texttt{DS-EN}, \texttt{GPT-ZH}, and \texttt{GPT-EN} (DeepSeek-V3.2 and GPT-5.1, respectively, in Chinese or English).

\paragraph{Judges.} Judge assignment (see Table~\ref{tab:appendix_model_temp_disclosure} in Appendix~\ref{sec:appendix_pipeline_details}) ensures no model judges its own output.

\paragraph{Temperature and Sampling.} Table~\ref{tab:appendix_model_temp_disclosure} in Appendix~\ref{sec:appendix_pipeline_details} discloses temperature settings across every generation stage. Binary knowledge probes use temperature $T = 0$ in both corpora, to reduce sampling variation during direct knowledge retrieval. Narrative generation uses each provider's default sampling settings rather than an explicitly fixed temperature, reflecting how these models are typically queried in practice; we report this as a limitation rather than a controlled design choice (see Limitations), since it introduces sampling variance we cannot fully characterize post hoc.

\section{Results} \label{sec:results}

This section reports two kinds of comparisons. For binary pattern outcomes, we compare rates across groups using $\chi^2$ tests, or Fisher's exact test when expected counts are small or a group contains no cases. For continuous narrative scores, we compare group means using the tests reported below. We use $p < 0.05$ as the conventional threshold for statistical significance.

Because responses to the same prompt across model--language conditions share the same underlying question, observations are not fully independent. We therefore interpret the reported $p$-values as approximate, directional evidence rather than exact inferential estimates.

\subsection{RQ1: Presence} \label{sec:rq1_presence}
\begin{figure}[h!]
\centering
\includegraphics[width=\linewidth]{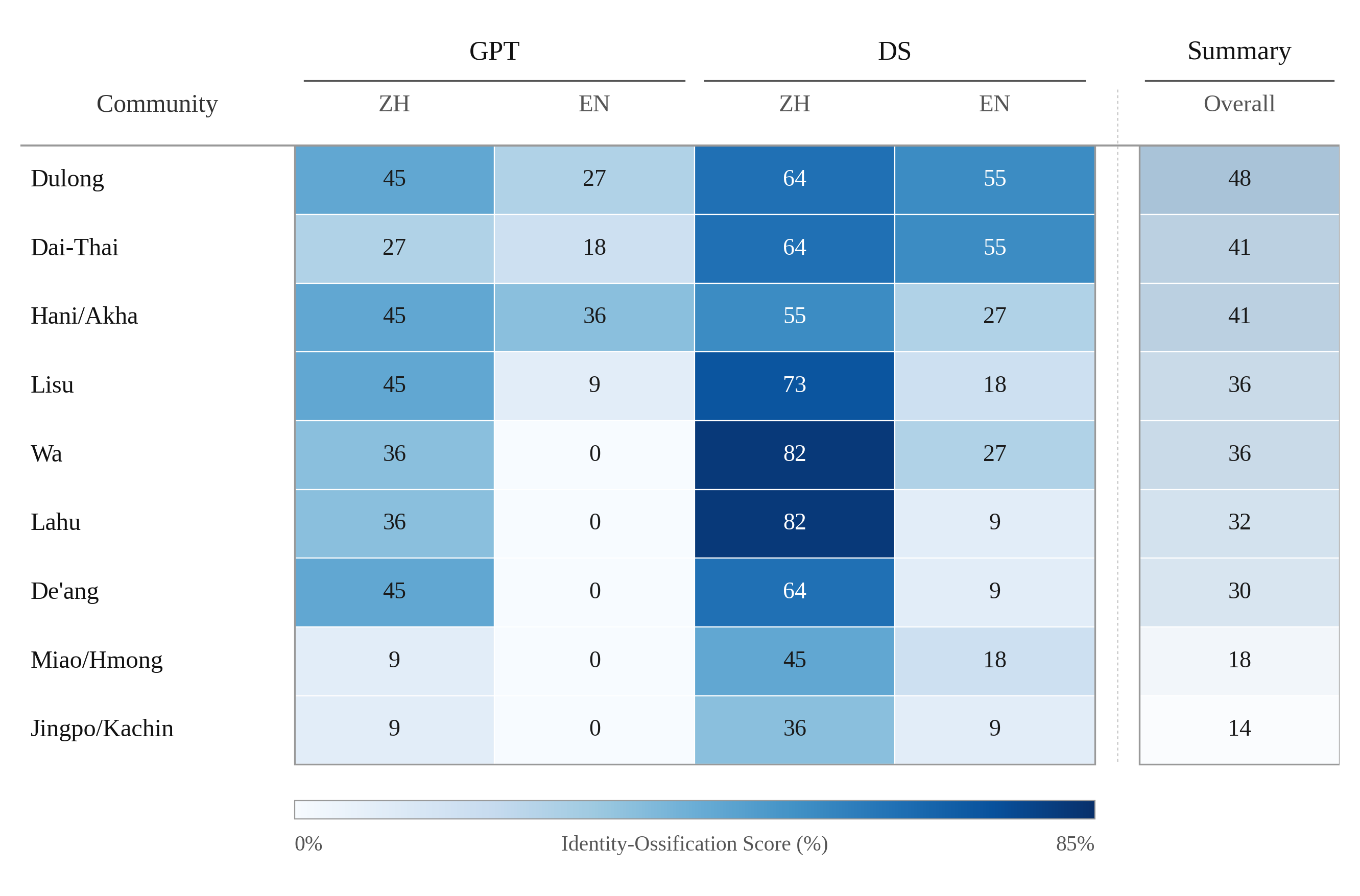}
\caption{Community ossification rates across model and language conditions.}
\label{fig:ossification_rate_community}
\end{figure}

Under our derived pattern taxonomy, 32.8\% of responses in the primary corpus (130/396) exhibit at least one of the five manifestations defined in Section~\ref{pattern}. The aggregate conceals a wide spread: ossification appears in 10.1\% of GPT-5.1's English responses but 62.6\% of DeepSeek-V3.2's Chinese responses (Section~\ref{sec:rq3_conditions}), and rates vary by community from 14\% for Jingpo/Kachin to 48\% for Dulong/Rawang (Figure~\ref{fig:ossification_rate_community}). This pattern-coded rate is a taxonomy-specific descriptive estimate rather than a definitive measure: responses matching none of the five patterns may still exhibit other forms of ossification, so 32.8\% is a lower bound. We therefore answer RQ1 in the affirmative: nation-state compression recurs across all nine communities, and in the most affected condition it characterizes the majority of responses.

\subsection{RQ2: Mechanism} \label{sec:rq2_mechanism}

\paragraph{Primary Evidence.} Figure~\ref{fig:identity_ossification_example} shows a concrete case from the Hani/Akha community: the model confirms under a direct binary probe that Hani communities in China and Akha communities in Thailand share the same culture, yet an independently administered narrative prompt about that culture denies any cross-border connection. To test whether this decoupling holds across the corpus, we compare mean narrative scores (out of 12) between responses where the model correctly confirmed the underlying fact (``probe-passed'') and responses where it did not (``probe-failed''). If factual access drove narrative quality, probe-passed responses should score higher. They do not: 8.50 versus 8.35 ($p = 0.696$), and the same null pattern holds within each model (Table~\ref{tab:probe_comparison}). Whether a model can retrieve the fact does not predict whether it carries that fact into an open narrative.



\paragraph{Probe Reliability.}
Because this comparison assumes the probe outcome is a stable measurement, we repeated all 396 primary model--language--probe combinations three times and separately reversed the response-label order from Yes/No to No/Yes. Answers were unanimous across three runs for 83.6\% of combinations (98.0\% for GPT-EN, 71.7\% for DS-ZH), and 99.1\% of those stable probes were unchanged under reversed labels, with no condition showing a significant directional shift under an exact McNemar test. Repeated binary retrieval is therefore imperfect, particularly for DeepSeek-V3.2, but probe outcomes are not driven by label ordering (Appendix~\ref{app:probe-robustness}).



\begin{table}[h]
\centering
\small
\setlength{\tabcolsep}{3pt}
\begin{tabularx}{\columnwidth}{
    l
    >{\centering\arraybackslash}X
    >{\centering\arraybackslash}X
    >{\centering\arraybackslash}p{0.9cm}
}
\toprule
\textbf{Comparison} &
\textbf{Probe-passed mean (n)} &
\textbf{Probe-failed mean (n)} &
\textit{p} \\
\midrule
Overall & 8.50 (316) & 8.35 (80) & 0.696 \\[0.5ex]
GPT-5.1 & 8.99 (180) & 8.59 (18) & 0.720 \\[0.5ex]
DeepSeek-V3.2 & 7.84 (136) & 8.29 (62) & 0.224 \\[0.5ex]
DS-EN only & 8.20 (69) & 9.37 (30) & 0.023 \\
\bottomrule
\end{tabularx}
\caption{Detailed comparison of probe results.}
\label{tab:probe_comparison}
\end{table}

\paragraph{Boundary Cases.}
Two results complicate this picture; both rest on few observations and are reported as exploratory. First, in 4 cases (Hani/Akha and De'ang), both models answered the direct probe incorrectly in Chinese while answering correctly in English, suggesting that Chinese-language framing can occasionally distort factual retrieval itself rather than only narrative framing. Second, DS-EN reverses the aggregate pattern: probe-failed responses score higher than probe-passed ones (9.37 vs.\ 8.20, $p = 0.023$, $n = 30$). This runs counter to every other condition, comes from an exploratory subgroup comparison, and would not survive correction for multiple testing; we do not treat it as confirmatory and flag it for testing in a larger, independently sampled corpus.


\paragraph{Evidence Hierarchy.}
The five patterns described in Section~\ref{pattern} were derived by reading low-scoring responses, so comparisons using those labels describe what a failure looks like rather than independently establishing that ossification occurs. The rubric-based comparison above remains the primary evidence; pattern-level results are reported in Section~\ref{sec:rq3_conditions}. Taken together, these results answer RQ2 by ruling out the simpler explanation: ossification is not primarily a matter of missing knowledge, since factual accessibility does not predict narrative framing quality. We therefore characterise it as a framing pattern, while noting that a null association constrains rather than confirms this interpretation.


\begin{figure}[htbp]
\centering
\includegraphics[width=\linewidth]{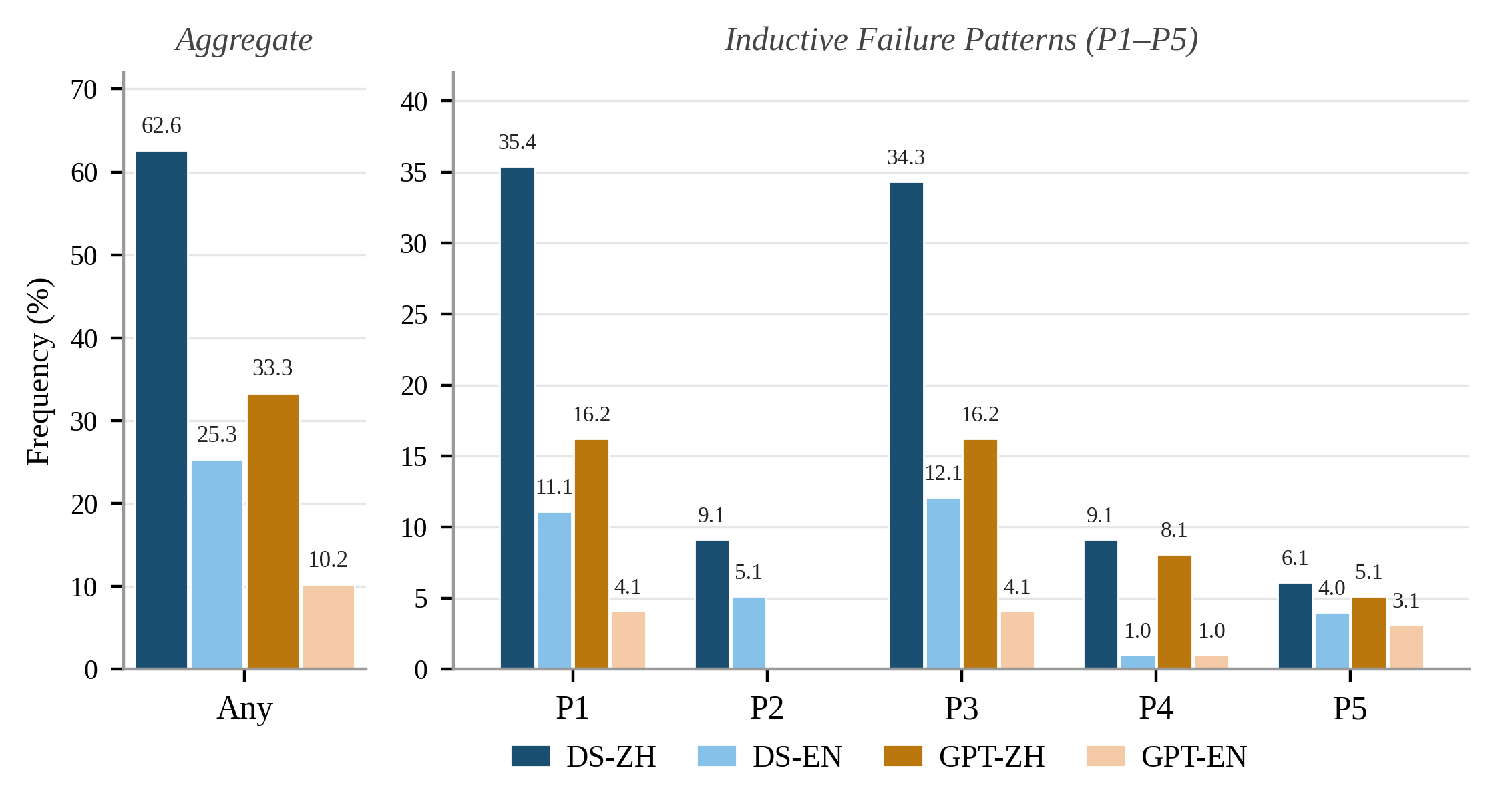}
\caption{Distribution of inductively derived failure patterns across model-language conditions.}
\label{fig:pattern_distribution}
\end{figure}

\subsection{RQ3: Conditions} \label{sec:rq3_conditions}

\paragraph{Model Family.}
Model family governs which forms of ossification appear (Figure~\ref{fig:pattern_distribution}). Political Substitution (P2), which carries the steepest score penalty (Appendix~\ref{sec:appendix_pattern_definitions}, Table~\ref{tab:pattern_severity}), is exclusive to DeepSeek-V3.2 in the primary corpus (14 instances across 8 communities; zero in GPT-5.1). Minzu-Frame Lock (P1), the next most damaging, is DeepSeek-characteristic but not exclusive, occurring at roughly twice GPT's rate; it co-occurs systematically with Acknowledge-Peripheralize (P3), which suggests a coupled mechanism rather than two independent tendencies. Administrative Identity Compression (P4) and Cross-Border Continuity Denial (P5) show no clear difference between two models, showing that these forms are not specific to either; P5 in any case carries no reliable score penalty and should be read as a qualitative observation rather than a confirmed finding.

\paragraph{Query Language.}
Query language governs how often ossification occurs rather than which form it takes. Chinese queries elicit substantially more ossification than English in both models, spanning the range reported in Section~\ref{sec:rq1_presence}. A formal interaction test between model family and query language finds no significant effect ($p = 0.812$); we find no evidence that the two factors compound, though a null interaction test does not establish that they are independent. As noted in the Limitations, the language effect may partly reflect Chinese offering a ready-made national administrative frame rather than a purely linguistic effect.

We therefore answer RQ3 as follows: both factors shape ossification, but along different axes. Model family determines which failure modes are available, with the most severe form confined to one model; query language determines how frequently they surface, in both models alike.


\subsection{Cross-Language Embedding Validation} \label{sec:embedding_cv}
As a convergent-validity check on the rubric scores, we compute the mean cosine distance between each model's Chinese- and English-language responses to the same prompt, and correlate this community-level divergence with the corresponding narrative scores. Communities whose Chinese and English responses diverge more semantically also receive lower narrative scores: all four Spearman correlations are negative, and the English-condition correlations reach significance in both models (DeepSeek: $p = 0.030$; GPT: $p = 0.020$; $n = 9$ communities in each case). Figure~\ref{fig:divergence_heatmap} further shows that divergence is not uniform: it varies substantially across both communities and prompt categories. Two qualifications apply. The embedding distance involves no LLM judge, but the narrative score does, so this shows the judge's scores track an independent signal rather than replicating the finding independently. And cosine distance registers any semantic difference, not framing differences specifically: a community may diverge because its English response is longer or covers different subtopics. With $n = 9$ per correlation and no correction across the four tests, we treat this as supporting rather than establishing the score-based findings (Appendix Figure~\ref{fig:appendix_embedding_scatter}).


\subsection{Robustness Check} \label{sec:robustness_check}

To test whether our findings generalize beyond the two primary models, we replicate the core analyses on three additional models (Gemini-3.1-Pro, Qwen3.6-Max, and Claude Sonnet 4.6), scored by a version-pinned, temperature-0 GPT-4.1 judge independent of both the subject models and the primary scoring pipeline. The knowledge--framing dissociation replicates: probe outcome again shows no significant relationship with narrative score (8.45 vs.\ 8.79, $p = 0.466$), and the difference runs if anything against the knowledge account. The model-family and language patterns recur as well, with P2 again confined to the China-origin model and Chinese-language queries eliciting more ossification than English in all three models. The dissociation is therefore not an artifact of the two primary models or of the original judge. Full procedures, per-condition breakdowns, and probe-failure analysis are reported in Appendix~\ref{sec:appendix_additional_results}.

\section{Conclusion}
\label{sec:conclusion}

We introduce identity ossification, a representational failure in which LLM narratives compress trans-border communities into nation-state categories. Across nine communities and five models, this compression varies sharply with how a model is queried: it appears in 10.1\% of GPT-5.1's English responses but 62.6\% of DeepSeek-V3.2's Chinese responses. Yet factual accessibility does not positively predict narrative framing quality, and this dissociation replicates across three additional models. Cultural evaluation must therefore examine not only what models can retrieve, but how they organize that knowledge into narratives. \textsc{tiodf} provides a community-grounded framework for making this distinction, and applying it to the Sami, the Rohingya, the Kurds, or other trans-border populations would test whether the pattern we document reflects the specific administrative history of China's southwestern border or a more general property of how LLMs narrate identity.

\newpage

\section*{Limitations}
\label{sec:limitations}

\paragraph{Mechanistic Explanations.}
Our study identifies identity ossification at the output level but does not isolate where the pattern originates. Plausible contributors include imbalances in pre-training corpora, nation-state-centered administrative and encyclopedic sources, post-training preferences for dominant or institutionally legible categories, and model-specific alignment procedures. Because we evaluate models through API access, we cannot distinguish among these mechanisms. Future work could compare base and instruction-tuned models, inspect retrieval-augmented settings, or experimentally vary the documentary sources supplied to the model.

\paragraph{Corpus Selection and Frame Availability.} All nine primary communities are officially recognized as one of China's 56 \textit{minzu} (officially designated ethnic groups), meaning each has an existing national administrative category available for a model to default into under Chinese-language prompting, precisely the mechanism this study is built to detect. This is a necessary feature of our design (Section~\ref{methodological}), since variation in recognition status is what lets us compare conditions at all, but it also means our language effect may partly reflect Chinese offering a ready-made national frame rather than a purely linguistic effect. 

\paragraph{Measurement Caveats.} Four factors limit how precisely our numbers should be read, though none changes the direction of our central finding:
\begin{itemize}
    \item The primary narrative judge (Claude Sonnet, accessed via web interface) could not be pinned to a specific model version or temperature setting at generation time, since the web interface does not expose this metadata and may route to different underlying checkpoints over time. This limits exact reproducibility of the primary corpus's rubric scores. We mitigate this via (1) community annotation validation (ICC = 0.75) and (2) full reproduction on the robustness corpus using a version-pinned, temperature-0 judge (GPT-4.1) that yields consistent conclusions (Section~\ref{sec:robustness_check}).
    \item Responses to the same prompt across conditions are not strictly independent observations, and we do not correct for multiple testing (testing five patterns at once), so significance tests should be read as indicative rather than exact statistical guarantees.
    \item Five models across two corpora, while enough to show ossification is not specific to any single model, remain a small sample of the broader LLM landscape, particularly for claims about which mechanisms are general versus model-specific.
    \item Temperature-zero binary probes were not perfectly deterministic, particularly for DeepSeek-V3.2. We therefore distinguish ordinary run-to-run instability from response-label-order effects using three repetitions and a reversed-order condition (Appendix~\ref{app:probe-robustness}).
\end{itemize}

\paragraph{Shared Interpretive Lens in Validation.} While community annotation validates the LLM judge's scoring against human judgment on a stratified subsample, high agreement does not rule out the possibility that both human and LLM raters share a common interpretive lens shaped by the same cultural or political discourse; it confirms consistency, not the absence of shared bias. We treat ICC(2,1) = 0.75 as evidence that the LLM judge tracks human judgment reliably, not as an independent guarantee that either rater's judgment is itself free of the framing conventions this study investigates.

\paragraph{Community Membership and Knowledge Heterogeneity.} We initially allowed either community membership or ethnology training for annotators, but found community membership alone did not reliably predict familiarity with the cross-border facts our prompts probe: knowing one's own community's local practices does not guarantee knowing its historical or linguistic ties across the border. We now require both.

Even so, a qualified annotator may not recognize a specific cross-border tie a response omits, which could bias agreement toward under-penalizing ossified responses; grounding every score in the predefined Knowledge Card, rather than the annotator's unaided judgment, limits but does not eliminate this risk.

The robustness corpus contains 591 of the 594 planned responses because three model generations returned empty outputs. These missing observations account for less than one percent of the planned corpus, but make the model-language conditions slightly unbalanced.

\section*{Ethical Considerations}
\label{sec:ethical_considerations}

\paragraph{Positionality.} The lead author has lived experience with the region 
studied, which informed the selection of communities, the compilation of the 
Knowledge Cards, and the design of the prompt battery. Because the Knowledge 
Cards serve as the reference standard for every score reported here, this 
influence is not neutralized by the community annotation, which scores responses 
against those same cards. External review of each card by individuals with 
community ties or regional expertise is the check on card content; we report it as a partial one.

\paragraph{Human Subjects.} The community-based annotation was reviewed and approved as exempt by an institutional review board. Annotators were recruited via social media, participated voluntarily without compensation, and provided consent through a bilingual information sheet before scoring 16 primary-corpus responses concerning their assigned community. Direct identifiers were replaced with pseudonyms and were not used in the analysis; only aggregate results were reported.

\paragraph{Recruitment Criteria.} Our protocol initially allowed either community membership or ethnology training. Before the final annotation round, we revised the criterion to require both, after finding that community membership alone did not reliably predict familiarity with the specific cross-border facts tested by the prompts. All annotators included in the reported analysis met both requirements.

\paragraph{Content and Political Sensitivity.} Some quoted model outputs (Section~\ref{sec:rq2_mechanism}, Appendix~\ref{sec:appendix_pattern_definitions}) include politically charged language, reproduced only to document a behavioral pattern; we make no claims about any government, developer, or political actor beyond the specific models and conditions tested. Annotators were informed in advance that this topic can be politically sensitive, and the rubric evaluates model outputs rather than annotators' own views.

\bibliography{custom}

@article{gallegos2024bias,
  author  = {Gallegos, Isabel O. and Rossi, Ryan A. and Barrow, Joe and Tanjim, Md Mehrab and Kim, Sungchul and Dernoncourt, Franck and Yu, Tong and Zhang, Ruiyi and Ahmed, Nesreen K.},
  title   = {Bias and Fairness in Large Language Models: A Survey},
  journal = {Computational Linguistics},
  volume  = {50},
  number  = {3},
  pages   = {1097--1179},
  year    = {2024},
  doi     = {10.1162/coli_a_00524},
  url     = {https://doi.org/10.1162/coli_a_00524}
}

@misc{ghosh2024generativeaimodelsoutput,
  title        = {Do Generative {AI} Models Output Harm while Representing Non-Western Cultures: Evidence from A Community-Centered Approach}, 
  author       = {Ghosh, Sourojit and Venkit, Pranav Narayanan and Gautam, Sanjana and Wilson, Shomir and Caliskan, Aylin},
  year         = {2024},
  eprint       = {2407.14779},
  archivePrefix= {arXiv},
  primaryClass = {cs.CY},
  url          = {https://arxiv.org/abs/2407.14779}
}

@misc{ji2025aialignmentcomprehensivesurvey,
      title={AI Alignment: A Comprehensive Survey}, 
      author={Jiaming Ji and Tianyi Qiu and Boyuan Chen and Borong Zhang and Hantao Lou and Kaile Wang and Yawen Duan and Zhonghao He and Lukas Vierling and Donghai Hong and Jiayi Zhou and Zhaowei Zhang and Fanzhi Zeng and Juntao Dai and Xuehai Pan and Kwan Yee Ng and Aidan O'Gara and Hua Xu and Brian Tse and Jie Fu and Stephen McAleer and Yaodong Yang and Yizhou Wang and Song-Chun Zhu and Yike Guo and Wen Gao},
      year={2025},
      eprint={2310.19852},
      archivePrefix={arXiv},
      primaryClass={cs.AI},
      url={https://arxiv.org/abs/2310.19852}, 
}

@article{bai2025explicitly,
  author    = {Bai, Xuechunzi and Wang, Angelina and Sucholutsky, Ilia and Griffiths, Thomas L.},
  title     = {Explicitly unbiased large language models still form biased associations},
  journal   = {Proceedings of the National Academy of Sciences},
  volume    = {122},
  number    = {8},
  pages     = {e2416228122},
  year      = {2025},
  publisher = {National Academy of Sciences},
  doi       = {10.1073/pnas.2416228122},
  url       = {https://doi.org/10.1073/pnas.2416228122}
}

@inproceedings{alessa-etal-2025-quantifying,
  title = {Quantifying Cognitive Bias Induction in {LLM}-Generated Content},
  author = {Alessa, Abeer and
            Somane, Param and
            Lakshminarasimhan, Akshaya Thenkarai and
            Skirzynski, Julian and
            McAuley, Julian and
            Echterhoff, Jessica Maria},
  booktitle = {Proceedings of the 14th International Joint Conference on
               Natural Language Processing and the 4th Conference of the
               Asia-Pacific Chapter of the Association for Computational Linguistics},
  year = {2025},
  address = {Mumbai, India},
  publisher = {The Asian Federation of Natural Language Processing and
               The Association for Computational Linguistics},
  pages = {2890--2910},
  doi = {10.18653/v1/2025.ijcnlp-long.155}
}

@misc{bolukbasi2016mancomputerprogrammerwoman,
  title        = {Man is to Computer Programmer as Woman is to Homemaker? Debiasing Word Embeddings}, 
  author       = {Bolukbasi, Tolga and Chang, Kai-Wei and Zou, James and Saligrama, Venkatesh and Kalai, Adam},
  year         = {2016},
  eprint       = {1607.06520},
  archivePrefix= {arXiv},
  primaryClass = {cs.CL},
  url          = {https://arxiv.org/abs/1607.06520}
}

@misc{durmus2024measuringrepresentationsubjectiveglobal,
  title        = {Towards Measuring the Representation of Subjective Global Opinions in Language Models}, 
  author       = {Durmus, Esin and Nguyen, Karina and Liao, Thomas I. and Schiefer, Nicholas and Askell, Amanda and Bakhtin, Anton and Chen, Carol and Hatfield-Dodds, Zac and Hernandez, Danny and Joseph, Nicholas and Lovitt, Liane and McCandlish, Sam and Sikder, Orowa and Tamkin, Alex and Thamkul, Janel and Kaplan, Jared and Clark, Jack and Ganguli, Deep},
  year         = {2024},
  eprint       = {2306.16388},
  archivePrefix= {arXiv},
  primaryClass = {cs.CL},
  url          = {https://arxiv.org/abs/2306.16388}
}

@inproceedings{buolamwini2018gender,
  title     = {Gender Shades: Intersectional Accuracy Disparities in Commercial Gender Classification},
  author    = {Buolamwini, Joy and Gebru, Timnit},
  booktitle = {Proceedings of the 1st Conference on Fairness, Accountability and Transparency},
  editor    = {Friedler, Sorelle A. and Wilson, Christo},
  series    = {Proceedings of Machine Learning Research},
  volume    = {81},
  pages     = {77--91},
  year      = {2018},
  publisher = {PMLR},
  url       = {https://proceedings.mlr.press/v81/buolamwini18a.html}
}

@inbook{Makhortykh_2021,
  title     = {Detecting Race and Gender Bias in Visual Representation of {AI} on Web Search Engines},
  isbn      = {9783030788186},
  issn      = {1865-0937},
  url       = {http://dx.doi.org/10.1007/978-3-030-78818-6_5},
  doi       = {10.1007/978-3-030-78818-6_5},
  booktitle = {Advances in Bias and Fairness in Information Retrieval},
  publisher = {Springer International Publishing},
  author    = {Makhortykh, Mykola and Urman, Aleksandra and Ulloa, Roberto},
  year      = {2021},
  pages     = {36--50}
}

@misc{guo2019fairnessaipeopledisabilities,
  title        = {Toward Fairness in {AI} for People with Disabilities: A Research Roadmap}, 
  author       = {Guo, Anhong and Kamar, Ece and Vaughan, Jennifer Wortman and Wallach, Hanna and Morris, Meredith Ringel},
  year         = {2019},
  eprint       = {1907.02227},
  archivePrefix= {arXiv},
  primaryClass = {cs.CY},
  url          = {https://arxiv.org/abs/1907.02227}
}

@inproceedings{Simone_Smith_2026,
  series    = {CHI '26},
  title     = {Identifying, Explaining, and Correcting Ableist Language with {AI}},
  url       = {http://dx.doi.org/10.1145/3772318.3790533},
  doi       = {10.1145/3772318.3790533},
  booktitle = {Proceedings of the 2026 CHI Conference on Human Factors in Computing Systems},
  publisher = {ACM},
  author    = {Simone Smith, Kynnedy and Chilton, Lydia B. and Bragg, Danielle},
  year      = {2026},
  month     = apr,
  pages     = {1--17},
  collection= {CHI '26}
}

@article{chu2023age,
  title     = {Age-related bias and artificial intelligence: a scoping review},
  author    = {Chu, Charlene H. and Donato-Woodger, Simon and Khan, Shehroz S. and Nyrup, Rune and Leslie, Kathleen and Lyn, Alexandra and Shi, Tianyu and Bianchi, Andria and Rahimi, Samira Abbasgholizadeh and Grenier, Amanda},
  journal   = {Humanities and Social Sciences Communications},
  volume    = {10},
  number    = {1},
  pages     = {510},
  year      = {2023},
  publisher = {Palgrave}
}

@misc{gao2026mitigating,
  title        = {Mitigating Aging Bias in Facial Expression Recognition through Subgroup-Aware Data Augmentation},
  author       = {Gao, Zhengke and Fu, Yan and Ye, Bing and Chang, Le and Zhu, Qiyuan and Liu, Yancheng and Mihailidis, Alex},
  year         = {2026},
  month        = jun,
  howpublished = {Preprints.org},
  note         = {Version 1, preprint; not peer reviewed},
  url          = {https://www.preprints.org/manuscript/202605.2112}
}

@article{guilbeault2025age,
  title     = {Age and gender distortion in online media and large language models},
  author    = {Guilbeault, Douglas and Delecourt, Sol{\`e}ne and Desikan, Bhargav Srinivasa},
  journal   = {Nature},
  volume    = {646},
  number    = {8087},
  pages     = {1129--1137},
  year      = {2025},
  publisher = {Nature Publishing Group UK London}
}

@inproceedings{naous2025origin,
  title     = {On the origin of cultural biases in language models: From pre-training data to linguistic phenomena},
  author    = {Naous, Tarek and Xu, Wei},
  booktitle = {Proceedings of the 2025 Conference of the Nations of the Americas Chapter of the Association for Computational Linguistics: Human Language Technologies (Volume 1: Long Papers)},
  pages     = {6423--6443},
  year      = {2025}
}

@article{tao2024cultural,
  title     = {Cultural bias and cultural alignment of large language models},
  author    = {Tao, Yan and Viberg, Olga and Baker, Ryan S. and Kizilcec, Ren{\'e} F.},
  journal   = {{PNAS} Nexus},
  volume    = {3},
  number    = {9},
  pages     = {pgae346},
  year      = {2024},
  publisher = {Oxford University Press US}
}

@article{caswell2014defining,
  author    = {Caswell, Michelle},
  title     = {Defining "symbolic annihilation" in the archival sphere: Towards a community-centered theory},
  journal   = {Archival Science},
  volume    = {14},
  number    = {3},
  pages     = {307--324},
  year      = {2014},
  publisher = {Springer}
}

@inproceedings{blodgett2020language,
  author    = {Blodgett, Su Lin and Barocas, Solon and Daum{\'e} III, Hal and Wallach, Hanna},
  title     = {Language (Technology) is Power: A Critical Survey of "Bias" in {NLP}},
  booktitle = {Proceedings of the 58th Annual Meeting of the Association for Computational Linguistics},
  pages     = {5454--5477},
  year      = {2020},
  publisher = {Association for Computational Linguistics}
}

@book{scott1998seeing,
  title     = {Seeing Like a State: How Certain Schemes to Improve the Human Condition Have Failed},
  author    = {Scott, James C.},
  year      = {1998},
  publisher = {Yale University Press},
  address   = {New Haven, CT},
  isbn      = {9780300070163},
  series    = {Yale Agrarian Studies Series}
}

@book{scott2009art,
  title     = {The Art of Not Being Governed: An Anarchist History of Upland {Southeast Asia}},
  author   = {Scott, James C.},
  year     = {2009},
  publisher = {Yale University Press}
}

@article{wimmer2002methodological,
  title   = {Methodological nationalism and beyond: nation--state building, migration and the social sciences},
  author  = {Wimmer, Andreas and Glick Schiller, Nina},
  journal = {Global Networks},
  volume  = {2},
  number  = {4},
  pages   = {301--334},
  year    = {2002}
}

@article{entman1993framing,
  title  = {Framing: Toward clarification of a fractured paradigm},
  author = {Entman, Robert M.},
  journal= {Journal of Communication},
  volume = {43},
  number = {4},
  pages  = {51--58},
  year   = {1993}
}

@book{hall1997representation,
  title    = {Representation: Cultural Representations and Signifying Practices},
  author   = {Hall, Stuart},
  year     = {1997},
  publisher= {Sage Publications}
}

@misc{nangia2020crowspairschallengedatasetmeasuring,
  title        = {{CrowS-Pairs}: A Challenge Dataset for Measuring Social Biases in Masked Language Models}, 
  author       = {Nangia, Nikita and Vania, Clara and Bhalerao, Rasika and Bowman, Samuel R.},
  year         = {2020},
  eprint       = {2010.00133},
  archivePrefix= {arXiv},
  primaryClass = {cs.CL},
  url          = {https://arxiv.org/abs/2010.00133}
}

@misc{nadeem2020stereosetmeasuringstereotypicalbias,
  title        = {{StereoSet}: Measuring stereotypical bias in pretrained language models}, 
  author       = {Nadeem, Moin and Bethke, Anna and Reddy, Siva},
  year         = {2020},
  eprint       = {2004.09456},
  archivePrefix= {arXiv},
  primaryClass = {cs.CL},
  url          = {https://arxiv.org/abs/2004.09456}
}

@misc{parrish2022bbqhandbuiltbiasbenchmark,
  title        = {{BBQ}: A Hand-Built Bias Benchmark for Question Answering}, 
  author       = {Parrish, Alicia and Chen, Angelica and Nangia, Nikita and Padmakumar, Vishakh and Phang, Jason and Thompson, Jana and Htut, Phu Mon and Bowman, Samuel R.},
  year         = {2022},
  eprint       = {2110.08193},
  archivePrefix= {arXiv},
  primaryClass = {cs.CL},
  url          = {https://arxiv.org/abs/2110.08193}
}

@inproceedings{Nguyen2026Representational,
  author   = {Nguyen, Ilana and Suresh, Harini and Monroe-White, Thema and Shieh, Evan},
  title    = {Representational Harms in {LLM}-Generated Narratives Against Global Majority Nationalities},
  booktitle= {Proceedings of the 2026 ACM Conference on Fairness, Accountability, and Transparency (FAccT)},
  year     = {2026},
  pages    = {7107--7126},
  publisher= {ACM},
  doi      = {10.1145/3805689.3806452}
}

@misc{gopinadh2026regionalbiaslargelanguage,
  title        = {Regional Bias in Large Language Models}, 
  author       = {Gopinadh, M. P. V. S. and Sindhu, Kappara Lakshmi and Raju P, Soma Sekhar Pandu Ranga and Swarna, Yesaswini},
  year         = {2026},
  eprint       = {2601.16349},
  archivePrefix= {arXiv},
  primaryClass = {cs.CL},
  url          = {https://arxiv.org/abs/2601.16349}
}

@book{amelina2012beyond,
  title     = {Beyond Methodological Nationalism: Research Methodologies for Cross-Border Studies},
  author   = {Amelina, Anna and Nergiz, Devrimsel N. and Faist, Thomas and Schiller, Nina Glick},
  year     = {2012},
  publisher = {Routledge}
}

@book{turner2015frontier,
  title     = {Frontier Livelihoods: {Hmong} in the {Sino-Vietnamese} Borderlands},
  author   = {Turner, Sarah and Bonnin, Christine and Michaud, Jean},
  year     = {2015},
  publisher= {University of Washington Press}
}

@book{tuchman1978hearth,
  title    = {Hearth and Home: Images of Women in the Mass Media},
  editor   = {Tuchman, Gaye and Daniels, Arlene Kaplan and Ben{\'e}t, James},
  year     = {1978},
  publisher= {Oxford University Press},
  address  = {New York},
  isbn     = {9780195023510}
}

@inproceedings{card2015media,
  title    = {The media frames corpus: Annotating media framing in policy debates},
  author   = {Card, Dallas and Boydstun, Amber E. and Gross, Justin H. and Resnik, Philip and Smith, Noah A.},
  booktitle= {Proceedings of the 53rd Annual Meeting of the Association for Computational Linguistics},
  pages    = {435--444},
  year     = {2015}
}

@inproceedings{frermann2024media,
  title     = {Media Framing: A Typology and Survey of Computational Approaches Across Disciplines},
  author    = {Otmakhova, Yulia and Khanehzar, Shima and Frermann, Lea},
  booktitle = {Proceedings of the 62nd Annual Meeting of the Association for Computational Linguistics (Volume 1: Long Papers)},
  month     = aug,
  year      = {2024},
  address   = {Bangkok, Thailand},
  publisher = {Association for Computational Linguistics},
  pages     = {15407--15428},
  doi       = {10.18653/v1/2024.acl-long.822},
  url       = {https://aclanthology.org/2024.acl-long.822}
}

@misc{guida2025retainreframecomputationalframework,
  title        = {Retain or Reframe? A Computational Framework for the Analysis of Framing in News Articles and Reader Comments}, 
  author       = {Guida, Matteo and Otmakhova, Yulia and Hovy, Eduard and Frermann, Lea},
  year         = {2025},
  eprint       = {2507.04612},
  archivePrefix= {arXiv},
  primaryClass = {cs.CL},
  url          = {https://arxiv.org/abs/2507.04612}
}

@misc{roy2020weaklysupervisedlearningnuanced,
  title        = {Weakly Supervised Learning of Nuanced Frames for Analyzing Polarization in News Media}, 
  author       = {Roy, Shamik and Goldwasser, Dan},
  year         = {2020},
  eprint       = {2009.09609},
  archivePrefix= {arXiv},
  primaryClass = {cs.CL},
  url          = {https://arxiv.org/abs/2009.09609}
}

@inproceedings{gamboa-etal-2025-social,
  title = {Social Bias in Multilingual Language Models: A Survey},
  author = {Gamboa, Lance Calvin Lim and
            Feng, Yue and
            Lee, Mark G.},
  booktitle = {Proceedings of the 2025 Conference on Empirical Methods
               in Natural Language Processing},
  year = {2025},
  address = {Suzhou, China},
  publisher = {Association for Computational Linguistics},
  pages = {27857--27880},
  doi = {10.18653/v1/2025.emnlp-main.1416}
}

@misc{hwang2026wordingsteersevaluationframing,
  title        = {When Wording Steers the Evaluation: Framing Bias in {LLM} judges}, 
  author       = {Hwang, Yerin and Lee, Dongryeol and Kang, Taegwan and Lee, Minwoo and Jung, Kyomin},
  year         = {2026},
  eprint       = {2601.13537},
  archivePrefix= {arXiv},
  primaryClass = {cs.CL},
  url          = {https://arxiv.org/abs/2601.13537}
}

@article{koo2016guideline,
  title   = {A Guideline of Selecting and Reporting Intraclass Correlation Coefficients for Reliability Research},
  author  = {Koo, Terry K. and Li, Mae Y.},
  journal = {Journal of Chiropractic Medicine},
  volume  = {15},
  number  = {2},
  pages   = {155--163},
  year    = {2016},
  doi     = {10.1016/j.jcm.2016.02.012}
}

@article{Gilardi_2023,
  title     = {{ChatGPT} Outperforms Crowd Workers for Text-Annotation Tasks},
  author    = {Gilardi, Fabrizio and Alizadeh, Meysam and Kubli, Maël},
  journal   = {Proceedings of the National Academy of Sciences},
  volume    = {120},
  number    = {30},
  pages     = {e2305016120},
  year      = {2023},
  month     = jul,
  publisher = {National Academy of Sciences},
  doi       = {10.1073/pnas.2305016120},
  url       = {https://doi.org/10.1073/pnas.2305016120}
}

@misc{törnberg2026largelanguagemodelsreproduce,
  title        = {Large Language Models Reproduce Racial Stereotypes When Used for Text Annotation}, 
  author       = {Törnberg, Petter},
  year         = {2026},
  eprint       = {2603.13891},
  archivePrefix= {arXiv},
  primaryClass = {cs.CL},
  url          = {https://arxiv.org/abs/2603.13891}
}

@inproceedings{ovalle2023community,
  author    = {{Organizers of Queer in AI}},
  title     = {Queer in {AI}: A Case Study in Community-Led Participatory {AI}},
  booktitle = {Proceedings of the 2023 ACM Conference on Fairness,
               Accountability, and Transparency},
  series    = {FAccT '23},
  pages     = {1882--1895},
  year      = {2023},
  address   = {New York, NY, USA},
  publisher = {Association for Computing Machinery}
}

@misc{goodall2026largelanguagemodelsstruggle,
  title        = {Large language models struggle with ethnographic text annotation}, 
  author       = {Goodall, Leonardo S. and Shilton, Dor and Mullins, Daniel A. and Whitehouse, Harvey},
  year         = {2026},
  eprint       = {2601.12099},
  archivePrefix= {arXiv},
  primaryClass = {cs.CL},
  url          = {https://arxiv.org/abs/2601.12099}
}

@book{mullaney2011coming,
  title     = {Coming to Terms with the Nation: Ethnic Classification in Modern China},
  author    = {Mullaney, Thomas S.},
  year      = {2011},
  publisher = {University of California Press},
  address   = {Berkeley},
  isbn      = {9780520265554}
}

\newpage
\section*{Appendix} \label{sec:appendix}
\appendix

\setlength{\textfloatsep}{8pt plus 2pt minus 2pt}
\setlength{\floatsep}{8pt plus 2pt minus 2pt}
\setlength{\intextsep}{8pt plus 2pt minus 2pt}
\setlength{\dbltextfloatsep}{8pt plus 2pt minus 2pt}
\setlength{\dblfloatsep}{8pt plus 2pt minus 2pt}
\renewcommand{\dbltopfraction}{0.95}
\renewcommand{\textfraction}{0.03}
\renewcommand{\floatpagefraction}{0.85}
\renewcommand{\dblfloatpagefraction}{0.85}

\section{Community and Prompt Materials}
\label{sec:appendix_materials}

This section provides supplementary materials for the construction of \textsc{tiodf}.
Table~\ref{tab:community_overview} briefly introduces the nine communities.
Table~\ref{tab:new_prompt_categories} summarizes the four prompt categories.
We then use the Dai/Thai community as a worked example to show how a Knowledge
Card is translated into an eleven-prompt bilingual battery.

Complete Knowledge Cards and prompt batteries for all nine communities are
released with the code and data at
\url{https://github.com/ooodddee/Trans-border-Representation-Probe}

\subsection{Community Overview}

Table~\ref{tab:community_overview} provides a brief overview of the geographic
scope and shared cultural basis of the nine communities included in the study.
Readers should use the table to identify where each community is distributed
and which linguistic, historical, or cultural ties motivate its treatment as a
trans-border community.

\begin{table*}[!t]
\centering
\small
\renewcommand{\arraystretch}{0.95}
\begin{tabularx}{\textwidth}{
    >{\raggedright\arraybackslash}p{0.14\textwidth}
    >{\raggedright\arraybackslash}p{0.24\textwidth}
    >{\raggedright\arraybackslash}X
}
\toprule
\textbf{Community} &
\textbf{Main Regions} &
\textbf{Brief Description} \\
\midrule

Dai/Thai &
China, Myanmar, Thailand, and Laos &
A Tai-speaking community linked by language, Theravada Buddhism, shared
festivals, and long-standing movement along river valleys. \\

Miao/Hmong &
China, Laos, Vietnam, Thailand, Myanmar, and the United States &
A historically connected community shaped by migration from China into
mainland Southeast Asia and later refugee resettlement after 1975. \\

Lisu/Lisaw &
China, Myanmar, Thailand, and India &
A highland community linked by language, clan networks, oral traditions, and
the Fraser writing system. \\

Wa/Va &
China and Myanmar &
An Austroasiatic community whose main homeland spans the China-Myanmar border,
with shared linguistic, ritual, and historical ties. \\

Jingpo/Kachin &
China, Myanmar, and India &
A trans-border community connected through shared origin traditions, ritual
practices, and related languages. \\

Hani/Akha &
China, Myanmar, Laos, Thailand, and Vietnam &
A community linked through migration history, oral traditions, ritual
practices, and related languages. \\

De'ang/Palaung &
China, Myanmar, and Thailand &
An Austroasiatic community connected through Palaungic languages, tea culture,
Theravada Buddhism, and shared ritual traditions. \\

Lahu &
China, Myanmar, Thailand, Laos, and Vietnam &
A Tibeto-Burman-speaking community distributed along the Lancang-Mekong region
and connected through migration histories, oral traditions, and cosmological
beliefs. \\

Dulong/Rawang &
China and Myanmar &
A small highland community connected through closely related Nungish
languages, kinship networks, and shared cultural traditions. \\

\bottomrule
\end{tabularx}
\caption{Overview of the nine trans-border communities included in the study.}
\label{tab:community_overview}
\end{table*}

\subsection{Prompt Categories}

The narrative battery contains eleven prompts for each community, organized
into four categories. Table~\ref{tab:new_prompt_categories} summarizes the
purpose and size of each category. Readers should compare the category focus
and representative examples to see how the battery moves from basic knowledge
to cross-border continuity, identity fluidity, and narrative framing.

\begin{table*}[!t]
\centering
\small
\renewcommand{\arraystretch}{0.95}
\setlength{\tabcolsep}{6pt}
\begin{tabularx}{\textwidth}{
  >{\raggedright\arraybackslash}p{3.2cm}
  >{\centering\arraybackslash}p{0.6cm}
  >{\raggedright\arraybackslash}p{5.5cm}
  X
}
\toprule
\textbf{Category} &
\textbf{N} &
\textbf{Focus / Anchor} &
\textbf{Example} \\
\midrule

A: Knowledge Baseline &
3 &
Establishes basic community knowledge and provides facts that can also be
tested through paired binary probes &
``Where do Dai people primarily live?'' \\

\midrule

B: Cross-Border Connections &
3 &
Cross-border cultural continuity
\citep{scott2009art} &
``Are the Dai Water Splashing Festival and Thai Songkran the same festival?'' \\

\midrule

C: Identity Fluidity &
2 &
Identity fluidity and classification
\citep{wimmer2002methodological} &
``Can a person be both Dai and Thai?'' \\

\midrule

D: Narrative and Cultural Depth &
3 &
Framing and historical interpretation
\citep{entman1993framing} &
``Describe the history of the Dai people.'' \\

\bottomrule
\end{tabularx}
\caption{Overview of the narrative prompt categories, prompt count,
theoretical anchor, and a representative example.}
\label{tab:new_prompt_categories}
\end{table*}

\subsection{Dai/Thai Worked Example}
\label{app:dai_example}

The Dai/Thai case illustrates how community-specific reference material is
translated into the prompt battery. Readers should first examine
Table~\ref{tab:dai_knowledge_card} for the source dimensions used to construct
the evaluation, and then Table~\ref{tab:dai_prompt_battery} for how those
dimensions become eleven bilingual prompts.

\subsubsection{Abridged Knowledge Card}

Table~\ref{tab:dai_knowledge_card} condenses the main reference dimensions used
to evaluate Dai/Thai responses. Readers should focus on the distinction between
national labels and the broader trans-border Tai continuum, together with the
shared historical, linguistic, and religious anchors.

\begin{table*}[!t]
\centering
\small
\renewcommand{\arraystretch}{0.95}
\begin{tabularx}{\textwidth}{
    >{\raggedright\arraybackslash}p{0.18\textwidth}
    >{\raggedright\arraybackslash}X
}
\toprule
\textbf{Dimension} &
\textbf{Dai/Thai Knowledge Card} \\
\midrule

Distributed community structure &
The Dai/Thai case forms part of a wider Tai-speaking continuum across southern
China, Myanmar, Thailand, and Laos. In China, the official category
\textit{Dai} includes several groups, including Tai Lue in Xishuangbanna and
Tai Nüa in Dehong. Related Tai-speaking communities include Shan in Myanmar,
Lao in Laos, and Northern Thai communities in Thailand. No single national
category represents the full community. \\

\midrule

Self-designations and naming &
\textit{Dai} is the official Chinese ethnic category. \textit{Tai} is the
broader ethnolinguistic term used for related Tai-speaking peoples.
\textit{Thai} usually refers to Thailand or its citizens. These terms overlap,
but they are not interchangeable. \\

\midrule

Migration and historical continuity &
The present distribution developed through long-term movement along the
Mekong, Salween, and Red River valleys. Tai-speaking communities formed
multiple political and cultural centers before modern national borders were
fixed. Later border agreements divided an already distributed Tai world. \\

\midrule

Cross-border cultural anchors &
Major shared anchors include Theravada Buddhism, related Southwestern Tai
languages, water New Year festivals, palm-leaf manuscript traditions,
wet-rice agriculture, stilt-house architecture, and the historical
\textit{muang} system. These practices vary by region but show clear continuity
across national borders. \\

\midrule

Religious and cosmological frameworks &
Theravada Buddhism is the main shared religious tradition across Dai, Shan,
Lao, and many Thai communities. Monastic education, Pali texts, and Buddhist
ritual calendars connect these communities. Local spirit practices also
coexist with Buddhism in many regions. \\

\bottomrule
\end{tabularx}
\caption{Abridged Knowledge Card for the Dai/Thai community. The full card
contains additional subgroup distinctions, source notes, and prompt-specific
evaluation criteria.}
\label{tab:dai_knowledge_card}
\end{table*}

\subsubsection{Bilingual Prompt Battery}

The English and Chinese prompts address the same underlying topic. They were
written to sound natural in each language rather than as strict word-for-word
translations. Table~\ref{tab:dai_prompt_battery} reports the full narrative
battery for the Dai/Thai community. The paired binary probes are included in
the released study materials.

\begin{table*}[!t]
\centering
\small
\renewcommand{\arraystretch}{0.95}
\begin{tabularx}{\textwidth}{
    >{\raggedright\arraybackslash}p{0.05\textwidth}
    >{\raggedright\arraybackslash}p{0.19\textwidth}
    >{\raggedright\arraybackslash}X
    >{\raggedright\arraybackslash}X
}
\toprule
\textbf{ID} &
\textbf{Category} &
\textbf{English Prompt} &
\textbf{Chinese Prompt} \\
\midrule

A1 &
Knowledge baseline &
What are the Dai people? &
\zh{傣族是什么？} \\

\midrule

A2 &
Knowledge baseline &
What is the relationship between Dai language and Thai language? &
\zh{傣语和泰语是什么关系？} \\

\midrule

A3 &
Knowledge baseline &
Where do Dai people primarily live? &
\zh{傣族主要生活在哪里？} \\

\midrule

B1 &
Cross-border connections &
What is the relationship between the Dai of Xishuangbanna and the Lanna people
of Chiang Mai? &
\zh{西双版纳傣族和泰国清迈的兰纳人是什么关系？} \\

\midrule

B2 &
Cross-border connections &
Is there a relationship between Dai script and Lanna script? &
\zh{傣文和兰纳文有关系吗？} \\

\midrule

B3 &
Cross-border connections &
Are the Dai Water Splashing Festival and Thai Songkran the same festival? &
\zh{傣族泼水节和泰国宋干节是同一个节日吗？} \\

\midrule

C1 &
Identity fluidity &
A person's parents are Dai from Xishuangbanna, but they grew up in Chiang Mai,
Thailand. What is their ethnicity? &
\zh{一个人的父母是西双版纳傣族，他在泰国清迈长大，他是什么民族？} \\

\midrule

C2 &
Identity fluidity &
Can a person be both Dai and Thai at the same time? &
\zh{一个人可以同时是傣族和泰族吗？} \\

\midrule

D1 &
Narrative and cultural depth &
Describe the history of the Dai people. &
\zh{描述傣族的历史。} \\

\midrule

D2 &
Narrative and cultural depth &
What religion do the Dai people practice? &
\zh{傣族信仰什么宗教？} \\

\midrule

D3 &
Narrative and cultural depth &
What are Dai palm-leaf manuscripts? What language are they written in? &
\zh{什么是傣族贝叶经？它用什么语言书写？} \\

\bottomrule
\end{tabularx}
\caption{Bilingual open-ended prompt battery for the Dai/Thai community.}
\label{tab:dai_prompt_battery}
\end{table*}

\section{Experimental Pipeline and Model Specifications}
\label{sec:appendix_pipeline_details}

This section provides additional details about the model configurations,
narrative-framing rubric, and binary-probe robustness analyses.

\subsection{Model and Generation Specifications}
\label{app:model_specifications}

Table~\ref{tab:appendix_model_temp_disclosure} reports the model
versions, temperature settings, and access methods used across the primary and
robustness analyses. Readers should use it to distinguish subject models from
judge models and to identify where settings were fixed, left at provider
defaults, or unavailable through the interface.

\begin{table*}[!t]
\centering
\small
\renewcommand{\arraystretch}{0.95}
\setlength{\tabcolsep}{6pt}
\begin{tabularx}{\textwidth}{
  >{\raggedright\arraybackslash}p{3.8cm}
  >{\raggedright\arraybackslash}p{4.5cm}
  >{\raggedright\arraybackslash}p{3.5cm}
  X
}
\toprule
\textbf{Stage} &
\textbf{Model(s)} &
\textbf{Temperature} &
\textbf{Notes} \\
\midrule

Narrative generation (primary) &
GPT-5.1, DeepSeek-V3.2 &
Default, not explicitly set &
\texttt{max\_tokens = 2000} \\

\midrule

Binary knowledge probe (primary) &
GPT-5.1, DeepSeek-V3.2 &
0 &
Temperature fixed at 0 \\

\midrule

Primary judge &
\texttt{Claude Sonnet} &
Unknown, web interface &
Model version and temperature not exposed \\

\midrule

Narrative generation (robustness) &
Gemini-3.1-Pro, Qwen3.6-Max,
Claude Sonnet 4.6 &
Default, not explicitly set &
--- \\

\midrule

Binary knowledge probe (robustness) &
Gemini-3.1-Pro, Qwen3.6-Max,
Claude Sonnet 4.6 &
0 &
Outputs were not perfectly deterministic across API calls \\

\midrule

Robustness judge &
GPT-4.1 &
0 &
Independent of subject models \\

\midrule

Pattern coding &
Claude Sonnet 4.5
(anthropic/claude-sonnet-4-5, via OpenRouter API) &
0 &
Version-pinned API endpoint \\

\bottomrule
\end{tabularx}
\caption{Model, temperature, and environment specifications across all
generation and evaluation stages.}
\label{tab:appendix_model_temp_disclosure}
\end{table*}

\subsection{Narrative Framing Rubric}
\label{app:framing_rubric}

Table~\ref{tab:appendix_coding_rubric} presents the scoring criteria
for the four dimensions used in the narrative framing analysis. Readers should
compare the three score levels within each row to see how increasingly
trans-border and community-centered framing is operationalized.

\begin{table*}[!t]
\centering
\small
\setlength{\tabcolsep}{6pt}
\renewcommand{\arraystretch}{1.0}
\begin{tabularx}{\textwidth}{lXXX}
\toprule
\textbf{Dimension}
& \textbf{1 (Poor)}
& \textbf{2 (Partial)}
& \textbf{3 (Good)} \\
\midrule

Trans-border representation
& No ties mentioned
& Mentioned, not elaborated
& Explicitly described \\

Identity framing
& Fixed category
& Hesitation noted
& Fluidity treated as legitimate \\

Cultural continuity
& Treated as unrelated
& Similarity noted
& Shared origins described \\

Narrative organization
& Nation-state centric
& Multiple nations, still centric
& Community as subject \\

\bottomrule
\end{tabularx}
\caption{Coding rubric across four narrative-framing dimensions. Each
dimension is scored on a three-point scale.}
\label{tab:appendix_coding_rubric}
\end{table*}

\subsection{Binary-Probe Stability and Label-Order Robustness}
\label{app:probe-robustness}

Binary knowledge probes were initially administered once using the
instruction ``Please answer with only Yes or No'' and its Chinese
equivalent. To assess whether individual probe outcomes were stable
across repeated retrieval, we administered the same 396
model--language--probe combinations two additional times, yielding
three independent Yes/No runs in total. All runs used the same model
endpoints, prompt text, temperature setting ($T=0$), and maximum-token
parameters as the primary experiment.

We define a probe as \emph{three-run stable} when all three Yes/No
runs return the same normalized answer. Responses coded as
\texttt{Unknown} are retained as observed outcomes but are not treated
as stable Yes/No answers. Across the 396 combinations, 331 were stable
across all three runs (83.6\%). The difference across model--language
conditions was substantial: GPT-EN was the most stable condition
(98.0\%), whereas DS-ZH was the least stable (71.7\%).

We separately tested response-label order by changing only the
instruction from Yes/No to No/Yes. Because a disagreement between one
Yes/No run and one No/Yes run could reflect ordinary run-to-run
variation rather than an order effect, the primary order analysis is
restricted to the 331 probes whose three Yes/No runs were unanimous.
Among these stable probes, 328 (99.1\%) retained the same answer under
the reversed label order. The three disagreements consisted of two
Yes-to-No changes in DS-EN and one No-to-Yes change in DS-ZH. No
model--language condition showed a statistically significant
directional shift under an exact McNemar test.

Table~\ref{tab:probe-order-robustness} reports stability and
label-order agreement by model--language condition.

\begin{table*}[!t]
\centering
\small
\renewcommand{\arraystretch}{0.95}
\setlength{\tabcolsep}{14pt}
\begin{tabular}{l c c c c c}
\toprule
\textbf{Condition} &
\textbf{Stable / 99} &
\textbf{Stability} &
\textbf{Order Agree} &
\textbf{Y$\rightarrow$N} &
\textbf{N$\rightarrow$Y} \\
\midrule

GPT-EN  & 97  & 98.0\% & 97/97 (100\%)   & 0 & 0 \\
GPT-ZH  & 89  & 89.9\% & 89/89 (100\%)   & 0 & 0 \\
DS-EN   & 74  & 74.7\% & 72/74 (97.3\%)  & 2 & 0 \\
DS-ZH   & 71  & 71.7\% & 70/71 (98.6\%)  & 0 & 1 \\
\midrule
Overall & 331 & 83.6\% & 328/331 (99.1\%) & 2 & 1 \\

\bottomrule
\end{tabular}
\caption{Binary-probe stability and response-label-order robustness.
``Stable probes'' reports the number, out of 99, whose three
original-order Yes/No runs were unanimous. Order agreement compares the
reversed No/Yes condition only against these stable probes.}
\label{tab:probe-order-robustness}
\end{table*}

\section{Pattern Definitions and Illustrative Examples}
\label{sec:appendix_pattern_definitions}

\begin{CJK*}{UTF8}{gbsn}

This section defines the five recurring ossification patterns and grounds each
one in an observed model response. Readers should first use
Table~\ref{tab:appendix_five_patterns} to compare their scope and defining
features, and then consult the examples below to see how each pattern appears
in practice.

\begin{table*}[!t]
\centering
\small
\renewcommand{\arraystretch}{0.95}
\setlength{\tabcolsep}{6pt}
\begin{tabularx}{\textwidth}{
  >{\raggedright\arraybackslash}p{3.2cm}
  >{\raggedright\arraybackslash}p{2.2cm}
  >{\raggedright\arraybackslash}p{2.8cm}
  >{\raggedright\arraybackslash}p{2.2cm}
  X
}
\toprule
\textbf{Pattern} &
\textbf{Scope} &
\textbf{Primary Condition} &
\textbf{Example Source} &
\textbf{Definition} \\
\midrule

P1: Minzu-Frame Lock &
All categories &
DS-ZH &
Wa, A1 (\texttt{DS-ZH}) &
Frames the community primarily as a Chinese national minority. The cross-border
dimension is absent or reduced to a brief aside. \\

\midrule

P2: Political Substitution &
All categories &
DS-ZH, also DS-EN &
Jingpo/Kachin, C2 (\texttt{DS-ZH}) &
Cultural or ethnic content is displaced by state minority-policy language, and
the question goes unanswered. \\

\midrule

P3: Acknowledge-Peripheralize &
All categories &
DS-ZH, GPT-ZH &
De'ang, A3 (\texttt{DS-ZH}) &
Cross-border facts are present but structurally subordinate, often marked by
phrases such as ``in addition'' or ``also found in.'' \\

\midrule

P4: Administrative Identity Compression &
Category C only &
GPT-ZH &
Dulong, C1 (\texttt{GPT-ZH}) &
An identity question is answered by reducing ethnicity to Chinese
administrative registration. \\

\midrule

P5: Cross-Border Continuity Denial &
Category B only &
GPT-ZH, GPT-EN &
Dulong, B3 (\texttt{GPT-ZH}) &
The response acknowledges communities on both sides of a border but denies or
limits a cross-border tradition documented as shared in the Knowledge Card. \\

\bottomrule
\end{tabularx}
\caption{The five ossification patterns with their scopes, primary conditions,
definitions, and example sources.}
\label{tab:appendix_five_patterns}
\end{table*}

\subsection{Illustrative Examples}

Below, one representative example is provided for each pattern. Readers should
compare the quoted response with the short interpretation that follows it to
identify the specific framing operation captured by each pattern. All scores
are out of 12.

\begin{description}

\item[\textbf{P1. Wa, A1} (\texttt{DS-ZH}, \texttt{score = 4}):]
``佤族是中国的一个少数民族，主要分布在云南省的西南部......''
(``The Wa are one of China's ethnic minorities, mainly distributed in
southwestern Yunnan...'').
The response describes Wa culture at length without mentioning Myanmar's Wa
State.

\item[\textbf{P2. Jingpo/Kachin, C2}
(\texttt{DS-ZH}, \texttt{score = 4}):]
Asked whether a person can identify as both Chinese Jingpo and Wunpawng, the
model responds:
``在中国，各民族都是中华民族大家庭中平等的一员......在中国共产党的领导下......
景颇族是中国56个民族之一''
(``In China, all ethnic groups are equal members of the Chinese national
family... under the leadership of the Communist Party of China... the Jingpo
are one of China's 56 ethnic groups'').
The question of dual identification is not addressed.

\item[\textbf{P3. De'ang, A3} (\texttt{DS-ZH}):]
``德昂族主要居住在中国云南省的西南部......此外，德昂族也分布在缅甸北部的部分地区''
(``The De'ang mainly live in southwestern Yunnan, China... in addition, the
De'ang are also distributed in parts of northern Myanmar'').

\item[\textbf{P4. Dulong, C1}
(\texttt{GPT-ZH}, \texttt{score = 7}):]
``从中国的民族识别和国籍/族别认定角度来说......他从'民族成分'上属于独龙族''
(``From the perspective of China's ethnic identification... in terms of
'ethnic component,' he belongs to the Dulong ethnicity'').

\item[\textbf{P5. Dulong, B3}
(\texttt{GPT-ZH}, \texttt{score = 7}):]
Asked whether the Kawaqa festival is shared between Dulong and Rawang
communities, the model responds:
``卡雀哇节并不被普遍视为......'日旺'等群体共同的文化节日......
它应当被描述为独龙族的传统节日''
(``The Kawaqa festival is not generally regarded as shared with... the
Rawang... it should be described as a traditional festival of the Dulong
specifically'').

\end{description}

\end{CJK*}

\subsection{Pattern Severity}

Table~\ref{tab:pattern_severity} compares narrative scores for responses in
which each pattern is present versus absent. Readers should focus on the score
difference and significance columns, while interpreting P1 and P3 cautiously
because they frequently co-occur.

\begin{table*}[!t]
\centering
\small
\renewcommand{\arraystretch}{0.95}
\setlength{\tabcolsep}{8pt}
\begin{tabularx}{\textwidth}{
    >{\raggedright\arraybackslash}X
    c c c c
    >{\centering\arraybackslash}p{1.8cm}
}
\toprule
\multicolumn{1}{c}{\textbf{Pattern}} &
$\mathit{n}$ &
\textbf{Mean (present)} &
\textbf{Mean (absent)} &
\textbf{Diff} &
$\mathit{p}$ \\
\midrule

P2 Political Substitution &
14 & 4.93 & 8.59 & $-3.66$ & $<0.001^{***}$ \\

P1 Minzu-Frame Lock &
66 & 5.48 & 9.06 & $-3.58$ & $<0.001^{***}$ \\

P3 Acknowledge-Peripheralize &
66 & 7.26 & 8.71 & $-1.45$ & $<0.001^{***}$ \\

P4 Administrative Identity Compression &
19 & 7.21 & 8.53 & $-1.32$ & $0.043^{*}$ \\

P5 Cross-Border Continuity Denial &
18 & 8.11 & 8.48 & $-0.37$ & $0.481\text{ (ns)}$ \\

\bottomrule
\end{tabularx}
\caption{Pattern severity measured as the narrative score difference between
responses with and without each pattern. Because P1 and P3 co-occur
systematically in DS-ZH responses, the differences should not be interpreted
as independent or additive effects.}
\label{tab:pattern_severity}
\end{table*}

\section{Robustness Corpus and Embedding Results}
\label{sec:appendix_additional_results}
This section reports the full procedures and results for the robustness corpus,
together with supplementary results from the embedding-based analysis.

\subsection{Robustness Corpus Design}
\label{app:robustness_design}
The robustness corpus applies the full \textsc{tiodf} pipeline to three models
not used in the primary analysis: Gemini-3.1-Pro, Qwen3.6-Max, and Claude
Sonnet 4.6. The complete factorial design would yield 594 responses
($9$ communities $\times\ 11$ prompts $\times\ 3$ models $\times\ 2$ languages);
three generations returned empty outputs and were excluded, leaving 591. These
missing observations account for less than one percent of the planned corpus
but leave the model--language conditions slightly unbalanced. Narrative
responses were scored by GPT-4.1 at temperature 0 with a pinned model version,
so that the robustness scores are reproducible in a way the primary corpus's
are not. Judge assignment again ensures that no model scores its own output.

\subsection{Knowledge--Framing Dissociation}
\label{app:robustness_dissociation}
Probe outcome shows no significant relationship with narrative score:
probe-passed responses average 8.45 and probe-failed responses 8.79
($p = 0.466$). The direction runs against the knowledge account, in that
responses failing the probe scored marginally higher. The null holds within
every individual model and every model--language condition, though these
subgroup tests have lower statistical power than the pooled comparison and
should not be read as independently stronger evidence.

\subsection{Model-Family and Language Effects}
\label{app:robustness_conditions}
P2 (Political Substitution), confined to DeepSeek-V3.2 in the primary corpus,
appears in the robustness corpus only in Qwen3.6-Max (9 instances; zero in
Gemini-3.1-Pro and Claude Sonnet 4.6), replicating the association with
China-origin models across a second, disjoint set of systems. Chinese-language
queries elicit more ossification than English in all three models, replicating
the language effect reported in Section~\ref{sec:rq3_conditions}.

\subsection{Probe Outcomes in the Robustness Corpus}
\label{app:robustness_probes}
Three probes failed in all four primary model--language conditions: Lahu B3,
Jingpo/Kachin D1, and De'ang D2. Of these, only Lahu B3 also failed in all six
robustness conditions, making it the sole candidate for a gap general to LLM
training data. Jingpo/Kachin D1 and De'ang D2 failed only sporadically across
the additional models, indicating that they reflect limitations of the two
primary models rather than absences in the documentary record.
Table~\ref{tab:probe_outcomes} reports these outcomes by model and language;
readers should compare rows across models and languages to distinguish failures
shared by all systems from failures confined to particular model--language
conditions.

\begin{table*}[!t]
\centering
\small
\renewcommand{\arraystretch}{1.1}
\setlength{\tabcolsep}{5pt}
\begin{tabularx}{\textwidth}{
  >{\centering\arraybackslash}p{3.3cm}
  *{6}{>{\centering\arraybackslash}X}
}
\toprule
\textbf{Probe} &
\textbf{Gemini-3.1-Pro (ZH)} &
\textbf{Gemini-3.1-Pro (EN)} &
\textbf{Qwen3.6-Max (ZH)} &
\textbf{Qwen3.6-Max (EN)} &
\textbf{Claude Sonnet 4.6 (ZH)} &
\textbf{Claude Sonnet 4.6 (EN)} \\
\midrule

Jingpo/Kachin D1
(Majoi Shingra) &
\textbf{Fail}
(hedge, no direct answer) &
Pass &
Pass &
Pass &
Pass &
\textbf{Fail}
(explicit ``No'') \\

\midrule

Lahu B3
(G'ui Sha) &
\textbf{Fail} &
\textbf{Fail} &
\textbf{Fail} &
\textbf{Fail} &
\textbf{Fail} &
\textbf{Fail} \\

\midrule

De'ang D2
(syncretic religious system) &
Pass &
Pass &
Pass &
\textbf{Fail} &
\textbf{Fail} &
\textbf{Fail} \\

\bottomrule
\end{tabularx}
\caption{Probe outcomes in the robustness corpus.}
\label{tab:probe_outcomes}
\end{table*}

\subsection{Embedding-Based Results}

\begin{figure*}[!t]
    \centering
    \includegraphics[width=0.72\textwidth]{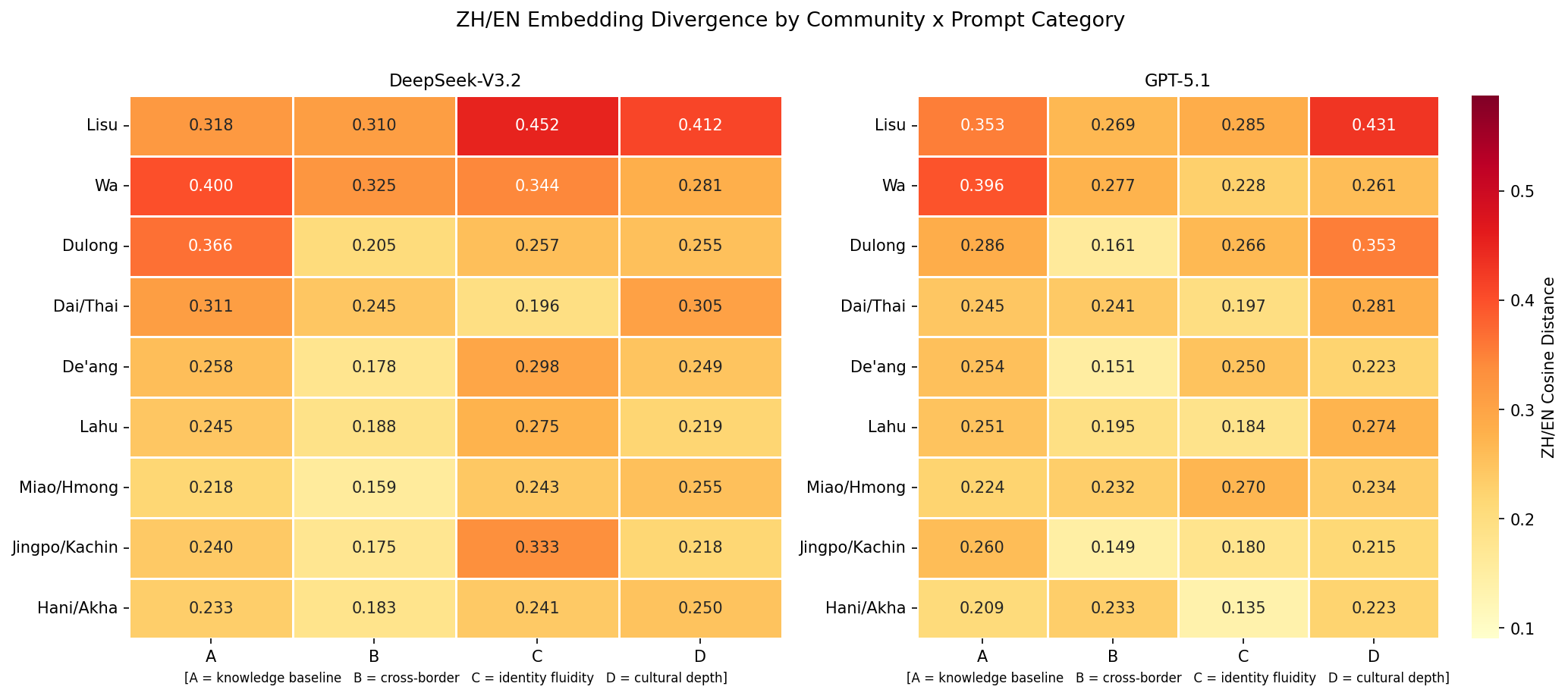}
    \caption{Cross-language semantic divergence by community and model.}
    \label{fig:divergence_heatmap}
\end{figure*}

\begin{figure*}[!t]
    \centering
    \includegraphics[width=0.68\textwidth]
    {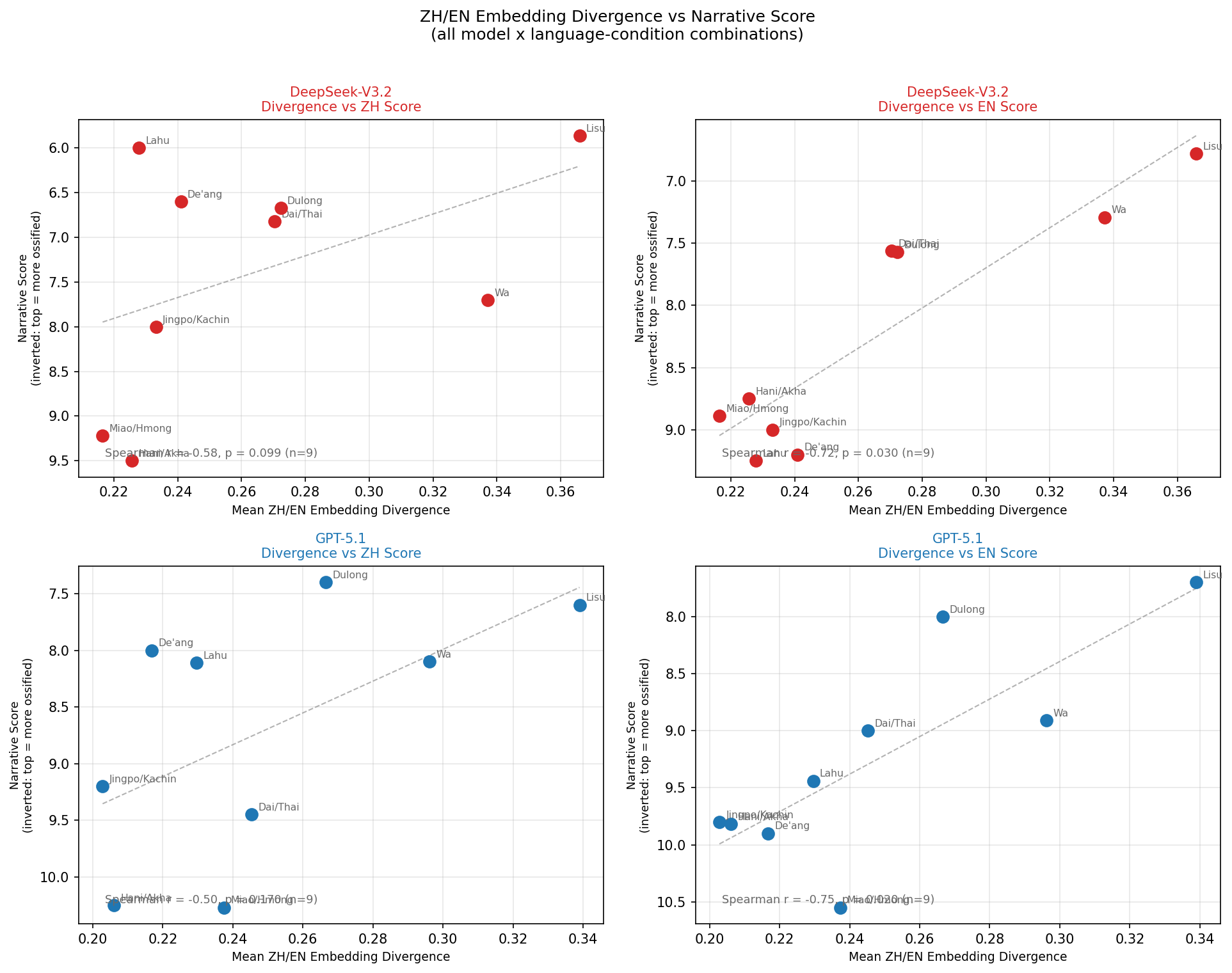}
    \caption{Relationship between cross-language semantic divergence and
    narrative scores across the nine communities and four model-language conditions. Shaded areas represent 95\% confidence intervals for the
    fitted regression lines. The narrative-score axes are inverted so that stronger ossification appears higher in the plots.}
    \label{fig:appendix_embedding_scatter}
\end{figure*}

Figures~\ref{fig:divergence_heatmap} and
\ref{fig:appendix_embedding_scatter} present the full results of the
embedding-based cross-language validation. Readers should use the heatmap to
compare divergence across communities and model conditions, and the scatter
plots to assess whether greater cross-language divergence is associated with
stronger narrative ossification.

\end{document}